\documentclass[%
 reprint,
 amsmath,amssymb,
 aps,
 floatfix,
]{revtex4-2}
\ifdefined\pdfminorversion
\fi

\usepackage{graphicx}%
\usepackage{dcolumn}%
\usepackage{bm}%
\usepackage{soul} %
\usepackage{xcolor}%
\usepackage{comment}

\graphicspath{{make_fig/}}

\begin{document}
\sethlcolor{red!20!white}%
\title{Delay-induced anomalous density--polarization relation in chiral active matter}%

\author{Dongrun Jian$^{1,2,3}$}
\author{Jie Su$^{2,3}$}%
\email{sj0410@ucas.ac.cn}
\author{Jun Wang$^{1}$}
\email{wangj@nju.edu.cn}
\author{Jin Wang$^{2,3,4}$}
\email{jin.wang.1@stonybrook.edu}
\affiliation{1.Department of Physics, Nanjing University, Nanjing, Jiangsu 210093, China\\
2.Center for Theoretical Interdisciplinary Sciences, Wenzhou Institute, University of Chinese Academy of Sciences, Wenzhou 325001, China\\
3.Zhejiang Key Laboratory of Soft Matter Biomedical Materials, Wenzhou Institute, University of Chinese Academy of Sciences, Wenzhou, Zhejiang 325000, China\\
4.Department of Chemistry and of Physics and Astronomy, State University of New York of Stony Brook, Stony Brook, New York 11794, USA}
\date{November 16, 2026}

\begin{abstract}
Aligning active matter can spontaneously exhibit diverse forms of ordered collective motion, and higher local density is generally expected to enhance local polarization. Meanwhile, delays in agents’ responses are widespread in real active systems, raising the question of how delayed alignment affects collective motion and, in particular, whether higher local density still strengthens polar order. Here, using a chiral active particle (CAP) model with delayed alignment, we uncover a counterintuitive anomalous density-polarization relation (DPR) within droplets: the local polarization first increases and then decreases with increasing local density.
Analysis of the deviation-angle dynamics reveals a delay-induced supercritical discrete Hopf bifurcation. Noise-sustained oscillations before the bifurcation and self-sustained oscillations afterward broaden the deviation-angle distribution and reduce polarization. Increasing the delay shifts the bifurcation to lower densities, enhancing the anomalous DPR in agreement with simulations. Using the nonequilibrium landscape-flux theory, we further identify the nonequilibrium flux as the dynamical origin of the anomalous DPR. Our findings not only establish time delay as a control knob for collective behavior in chiral active matter, but also offer a useful analytical perspective on anomalous density–order relations in other nonequilibrium systems exhibiting collective order.
\end{abstract}

\maketitle

Active matter systems consist of agents that consume energy to generate directed motion~\cite{Ramaswamymechanics2010}.
They range from bacterial suspensions and bird flocks~\cite{DiLuzioEscherichia2005,LingBehavioural2019} to artificial systems such as Janus particles and Quincke rollers~\cite{HowseSelfmotile2007,ReyesGarzaMagnetic2023}.
These systems exhibit diverse collective phenomena, including flocking, clustering, and active turbulence~\cite{BeccoExperimental2006,ButtinoniDynamical2013,ZhouCollective2025}.
In aligning active matter, local interactions tend to align the orientations of neighboring agents. By bringing more neighbors within alignment range, higher density can favor polar order (e.g., in Quincke roller bands and mobbing jackdaw flocks~\cite{BricardEmergence2013,LingBehavioural2019}). The Vicsek model \cite{VicsekNovel1995} provides a minimal description of polar alignment. It has been extensively studied \cite{GregoireOnset2004, ChateCollective2008, IhleInvasionwaveinduced2013, CaussinEmergent2014, KyriakopoulosClustering2019} and has spawned many variants \cite{ChateSimple2006, MInteraction2008, ChenForeaft2017}.
Among these, Liebchen et al. \cite{LiebchenCollective2017} proposed a chiral variant of the Vicsek model, called the chiral active particle (CAP) model, which describes active matter systems with kinetic chirality commonly found in nature \cite{KaiserVortex2013, FranquelimRevolving2017, ChenWeak2017, PatraCollective2022}.
The introduction of chirality in the CAP model not only generates unique dynamic and thermodynamic behaviors \cite{VentejouSusceptibility2021, JianUncovering2025, LiebchenChiral2022} but also gives rise to rich patterns such as micro-flocks and macro-droplets \cite{LiebchenCollective2017, KrukTraveling2020}.
As we show below, local polarization increases monotonically with local density within droplets of the CAP model. We refer to this monotonic relation as the \textit{normal density--polarization relation} (DPR).

In most Vicsek-like models (including the CAP model), an agent receives directional signals from its neighbors and changes its own direction simultaneously.
In real-world systems, however, agents require time to respond due to the finite speeds of information transfer, processing, and actuation, which naturally introduce delays \cite{AttanasiInformation2014}.
Remarkably, time delays have significant effects on active matter systems, such as inducing chirality \cite{WangSpontaneous2023} and forming collective patterns \cite{ChenActive2023} in active Brownian particles with delayed attractions.
Moreover, in flocking models \cite{HolubecFeedback2026} such as the Vicsek model \cite{PiwowarczykInfluence2019, HolubecFinitesize2021}, time delays provide an effective means to switch between dynamical states, with short delays promoting order and long delays promoting disorder.
These effects of delay on collective order motivate us to ask how delayed alignment reshapes the DPR within ordered droplets and what mechanisms underlie this change.

In this work, by incorporating delayed alignment into the CAP model, we develop a delayed CAP model and uncover a counterintuitive behavior within the formed droplets: local polarization first rises and then falls as local density increases, leaving the densest core less polarized than the periphery, contradicting the normal DPR in the delay-free CAP model.
We term this counterintuitive behavior the \textit{anomalous density--polarization relation}.
Focusing on this behavior, we derive the dynamical equation for the deviation angle of particles inside droplets and analyze it via bifurcation theory \cite{StrogatzNonlinear2024}. We identify a supercritical discrete Hopf bifurcation in the deviation-angle dynamics, producing a periodic orbit with self-sustained oscillations above a critical density. Even below this density, noise-sustained oscillations occur around the stable fixed point. Both oscillations broaden the deviation-angle distribution and reduce polarization, accounting for the anomalous DPR.
As the delay increases, the critical density decreases, shifting the bifurcation to lower densities, thereby making the anomalous DPR more pronounced, in agreement with the simulation results.
Moreover, using the nonequilibrium landscape-flux theory \cite{WangPotential2008,WangLandscape2015,FangNonequilibrium2019,SuDynamical2024,YanMechanism2025,JianUncovering2025}, we identify the nonequilibrium flux and its density derivative as the dynamical origin of the anomalous DPR and the signature of the bifurcation, respectively.
In addition, we also explore the influence of the alignment strength and chirality strength on the anomalous DPR.

\paragraph*{Model \lowercase{and} Methods.--}
In the delayed CAP model, each chiral active particle experiences an alignment force after a delay $\tau$ following the reception of neighbor signals and the computation of the alignment force. To be more precise, the dynamics of the $i$-th particle is governed by
\begin{equation}
	\label{eq:continued-cap}
	\begin{aligned}
		\dot{\mathbf{r}}_i(t)&= v \mathbf{n}_i(t),\\
		\dot{\theta}_i(t)&=\omega+
		\frac{K}{\pi R^2}\sum_{j\in N_i}\sin{(\theta_j(t-\tau)-\theta_i(t-\tau))}+\sqrt{2\eta}\zeta_i(t)
	\end{aligned}
\end{equation}
where $\mathbf{r}_i$, $v$, $\theta_i$, and $\mathbf{n}_i(t)=(\cos\theta_i(t),\sin\theta_i(t))$ are respectively the position vector, the self-propulsion speed, the angle, and the moving direction of the $i$-th particle.
The angular equation comprises chirality, alignment, and noise as its three terms.
$\omega$ is the chirality strength, representing the intrinsic rotation frequency of the particle. In the alignment term, $K$ and $R$ are the alignment strength and radius, respectively. $N_i$ denotes the set of all particles in the neighborhood of particle $i$ (the region within $R$), including particle $i$ itself.
$\eta$ is the noise intensity, and $\zeta_i(t)$ is the standard Gaussian white noise satisfying $\langle\zeta_i(t)\zeta_{i^\prime}(t^{\prime})\rangle=\delta_{ii^\prime}\delta(t-t^{\prime})$.
$\tau$ denotes the delay time. For $\tau=0$, the delay effect disappears, and the model reduces to the original CAP model.

In our simulation, we use $R$ and $R$/$v$ as basic units for length and time, respectively.
We discretize the equations \eqref{eq:continued-cap} by using the Euler integration method and perform numerical simulations with $N=32000$ particles placed in a two-dimensional box of size $L=80$ with periodic boundary conditions, corresponding to the global density $\rho_0=N/L^2=5$. We set $R=1$, $v=1$, $\eta=1.5$, and $\Delta t=0.01$, where $\Delta t$ is the time step. We set $\omega=0.5$ and $K=2$, if not otherwise stated.
The initial construction of the delayed CAP system is random.
All sampled data are collected after the system reaches steady state ($10^7$ time steps).

\paragraph*{Delay-induced anomalous DPR in ordered droplets.--}
To quantitatively describe the anomalous DPR, we introduce two parameters: the local density ($\rho_L=n/A$) and the local polarization ($\phi_L=\left|\frac{1}{n}\sum_{k\in A}e^{\mathrm{i}\theta_k}\right|$), where $A$ is the area of a local grid cell, and $n$ is the number of particles in $A$. Accordingly, regions with $\rho_L>\rho_0$ and $\rho_L<\rho_0$ correspond to the ordered droplet and the disordered gas phase, respectively.

\begin{figure}[!t]
	\centering
	\includegraphics[width=1\linewidth]{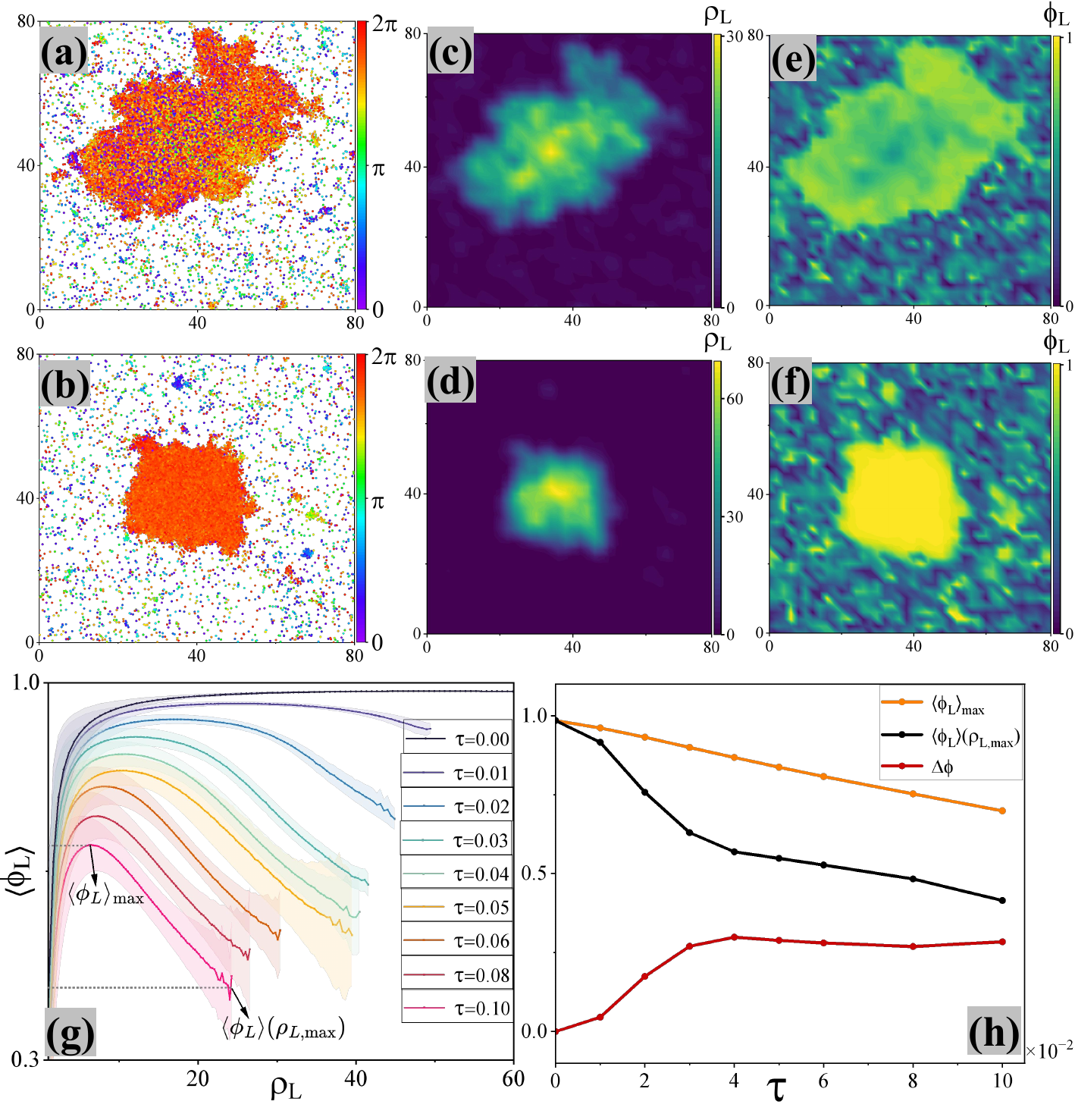}
	\caption{Anomalous density--polarization relation (DPR) induced by time delay.
	(a--f) Spatial distributions of the ordered droplet phase.
	(a,c,e) Snapshot, local density $\rho_L$, and local polarization $\phi_L$, respectively, for the delayed CAP system ($\tau=0.05$).
	(b,d,f) Snapshot, $\rho_L$, and $\phi_L$ for the delay-free CAP system.
	(g) Mean local polarization $\langle\phi_L\rangle$ vs $\rho_L$ at various $\tau$.
	(h) $\langle\phi_L\rangle_{\max}$ (orange), $\langle\phi_L\rangle(\rho_{L,\max})$ (black), and $\Delta\phi=\langle\phi_L\rangle_{\max}-\langle\phi_L\rangle(\rho_{L,\max})$ (red) as functions of $\tau$.}
	\label{fig:fig1}
\end{figure}

For the original and delayed CAP systems under our parameter setting, the ordered droplet phase can spontaneously emerge, as shown in Fig.~\ref{fig:fig1}.
In both systems, we observe that $\rho_L$ is higher in regions closer to the droplet core [Fig.~\ref{fig:fig1}(c,d)]. However, the spatial distribution of $\phi_L$ differs between the two systems. In the undelayed system [Fig.~\ref{fig:fig1}(f)], regions with higher $\rho_L$ near the core also exhibit higher $\phi_L$. In contrast, in the delayed system [Fig.~\ref{fig:fig1}(e)], we observe a novel behavior: the region with the highest $\rho_L$ at the core actually shows a lower $\phi_L$ than the peripheral regions.

For clarity in describing the anomalous DPR, we plot the mean local polarization $\langle\phi_L\rangle$ against $\rho_L$ for various $\tau$ in Fig.~\ref{fig:fig1}(g). 
In the absence of delay ($\tau=0$), $\langle\phi_L\rangle$ increases monotonically with $\rho_L$. 
This is consistent with the intuitive picture that a higher local density provides more neighbors, which strengthens the alignment force and thus enhances the local polarization.
However, for nonzero $\tau$, the anomalous DPR is induced: $\langle\phi_L\rangle$ first increases and then decreases with increasing $\rho_L$, exhibiting a non-monotonic dependence. As $\tau$ increases, $\langle\phi_L\rangle$ decreases globally, and its peak shifts to lower $\rho_L$, implying a more pronounced anomalous DPR.

To quantify the degree of the DPR anomaly, we define $\Delta\phi = \langle\phi_L\rangle_{\max} - \langle\phi_L\rangle(\rho_{L,\max})$, which measures the polarization deficit of the densest region relative to the most polarized region within the droplet. Here, $\langle\phi_L\rangle_{\max}$ is the maximum mean local polarization and $\rho_{L,\max}$ is the maximal local density. Thus, $\Delta\phi=0$ corresponds to a normal monotonic DPR, whereas a larger $\Delta\phi$ indicates a more pronounced anomalous DPR.
Fig.~\ref{fig:fig1}(h) summarizes the dependence of $\langle\phi_L\rangle_{\max}$ (orange), $\langle\phi_L\rangle(\rho_{L,\max})$ (black), and $\Delta\phi$ (red) on $\tau$. 
The quantity $\langle\phi_L\rangle_{\max}$ decreases linearly with increasing $\tau$, indicating that the delay reduces the overall polarization inside the droplet. 
Meanwhile, $\langle\phi_L\rangle(\rho_{L,\max})$ decreases more rapidly than $\langle\phi_L\rangle_{\max}$ does, but its decline slows down at larger $\tau$. 
Consequently, $\Delta\phi$ increases with $\tau$ first, and tends to saturate at large delays, confirming that the delay enhances the anomalous DPR in general.

\paragraph*{Dynamical bifurcation.--}

With the neighborhood quantities defined, the alignment term in the angular dynamics of Eq.~\eqref{eq:continued-cap}, inspired by the Kuramoto model \cite{Acebronkuramoto2005}, takes the form:
$K\rho_{N}^i(t-\tau)\phi_{N}^i(t-\tau)\sin(\alpha_{i}(t-\tau)-\theta_i(t-\tau))$,
where 
$\rho_{N}^i:=\frac{|N_i|}{\pi R^2}$, $\phi_N^i := \bigl|\frac{1}{|N_i|}\sum_{j\in N_i} e^{\mathrm{i}\theta_j}\bigr|$, and $\alpha_i := \operatorname{Arg}\bigl(\sum_{j\in N_i} e^{\mathrm{i}\theta_j}\bigr)$ 
are the density, polarization, and average angle of CAPs in the $i$-th particle's neighborhood, respectively (see Supplemental Material (SM) for details). 
Inside the steady-state droplet, we define the deviation angle $\psi_i(t):=\theta_i(t)-\alpha_i(t)$, and derive its evolution from Eq.~\eqref{eq:continued-cap} under the steady-state approximation (see SM for details):
\begin{equation}
\label{eq:SStateApprox2}
\dot{\psi}_i(t)= -\beta_i\sin(\psi_i(t-\tau))+\sqrt{2\eta}\zeta_i(t),
\end{equation}
where $\beta_i:=K\rho_{N}^i\phi_{N}^i$ is the effective alignment strength of the neighborhood ($\rho_{N}^i$ and $\phi_{N}^i$ are time-independent under our approximation explained in SM).
Thus, the original multi-particle coupled Eq.~\eqref{eq:continued-cap} is reduced to an equivalent single-particle equation in the neighborhood, Eq.~\eqref{eq:SStateApprox2}. Regardless of a particle's position within the droplet, as long as it experiences the same $\rho_N$ and $\phi_N$, the evolution of its deviation angle $\psi$ follows Eq.~\eqref{eq:SStateApprox2}. 

To reveal the dynamical mechanism of anomalous DPR, we consider the deterministic part of Eq.~\eqref{eq:SStateApprox2}. Discretizing it in time yields a single-variable delay-difference equation:
\begin{equation}
\label{eq:determ-dde}
\psi(t+\Delta t)-\psi(t)=-\beta \Delta t\sin(\psi(t-\tau)).
\end{equation}
Through a standard bifurcation analysis \cite{StrogatzNonlinear2024}, we characterize the dynamics of Eq.~\eqref{eq:determ-dde} (see SM for details). For $\tau>0$, at the critical point $\beta_c:=\frac{2}{\Delta t}\sin(\frac{\pi\Delta t}{2(2\tau+\Delta t)})$, the system undergoes a supercritical discrete Hopf bifurcation. In the limit $\Delta t\to 0$, $\beta_c\to \pi/(2\tau)$. 
For $0<\beta<\beta_c$, the fixed point $\psi^*=0$ is locally asymptotically stable. 
For $\beta>\beta_c$, the fixed point loses stability and a periodic orbit emerges, corresponding to self-sustained oscillations.
Linearizing the noisy version of Eq.~\eqref{eq:determ-dde} around $\psi^*$ predicts a polarization maximum at $\beta_s$ (see SM for details). As $\Delta t\to0$, $\beta_s/\beta_c$ approaches a constant ($\approx0.47$).
For $\beta_s<\beta<\beta_c$, noise continuously excites the damped oscillatory mode that loses stability at $\beta_c$, sustaining quasicycles~\cite{LugoQuasicycles2008,McKanePredator2005}. As $\beta$ increases, weaker damping amplifies the response to noise, enhancing fluctuations in $\psi$ and reducing polarization (see Fig. S2 in SM).
In short, the dynamics of $\psi$ is determined by $\beta$. 
Over the simulated density range, $\beta$ increases monotonically with $\rho_N$ (see Fig. S3 in SM), giving a one-to-one correspondence. We thus define the optimal and critical densities $\rho_s$ and $\rho_c$ as the densities corresponding to $\beta_s$ and $\beta_c$, respectively.

\begin{figure}[!t]
	\centering
	\includegraphics[width=1\linewidth]{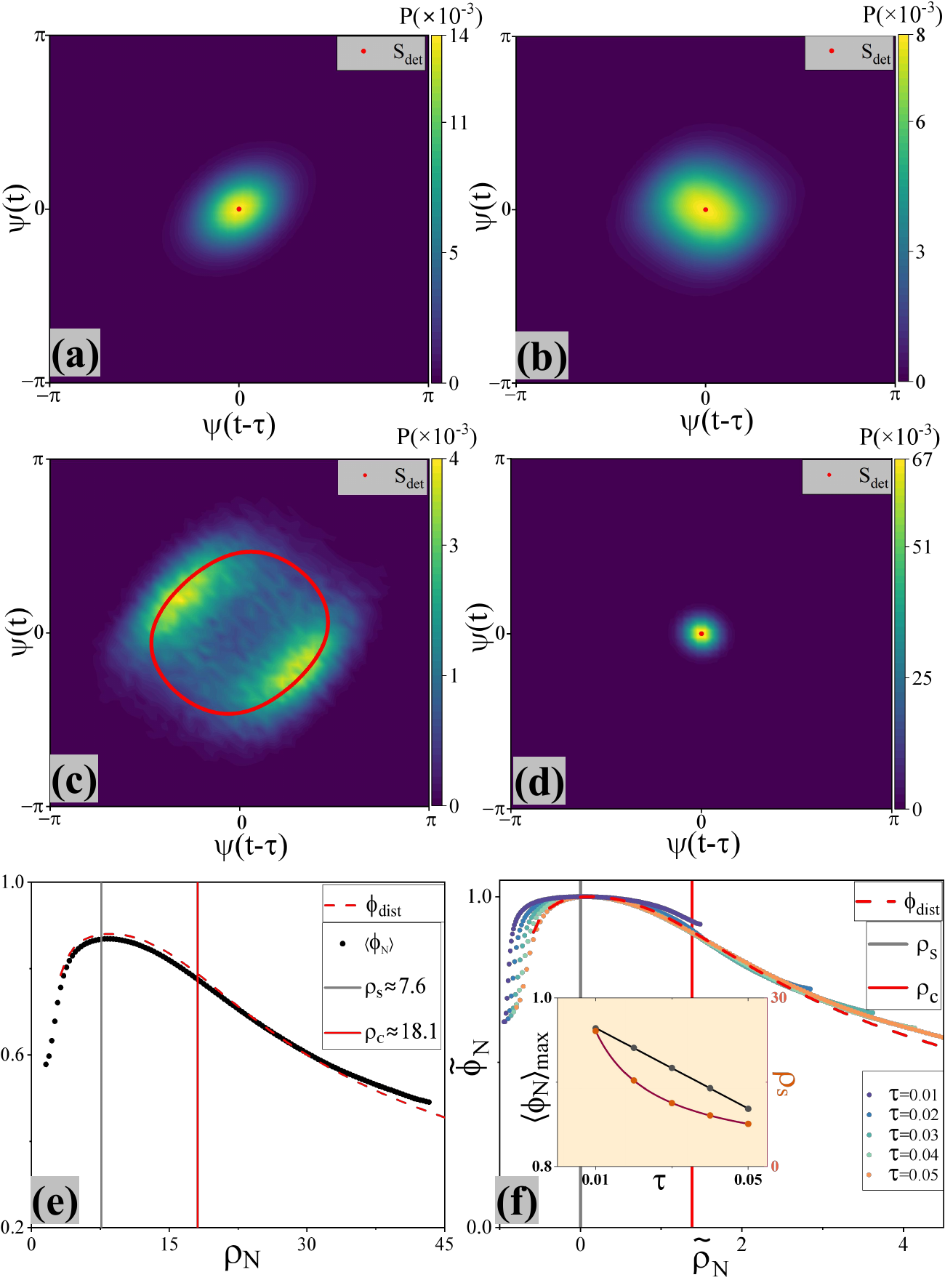}
	\caption{Bifurcation, theoretical polarization curves, and scaled master curves for the anomalous DPR.
	(a--d) Probability distribution $P$ (color map) and deterministic solution $S_\mathrm{det}$ (red) in the $(\psi(t-\tau),\psi(t))$ phase plane, for neighborhood densities (a) $\rho_N=9$, (b) $\rho_N=17$, and (c) $\rho_N=30$ with delay $\tau=0.05$, and (d) $\rho_N=30$ without delay.
	(e) Theoretical polarization $\phi_\mathrm{dist}$ and simulated polarization $\langle\phi_N\rangle$ versus $\rho_N$ for $\tau=0.05$. (f) Scaled polarization curves for different $\tau$ in the scaled coordinates $\tilde{\rho}_N$ and $\tilde{\phi}_N$.
	Inset of (f): the maximum polarization $\langle\phi_N\rangle_{\max}$ (black dots) and the optimal density $\rho_s$ (orange dots) as functions of $\tau$; solid lines are fits.
	In (e, f), the gray and red vertical lines mark the optimal density $\rho_s$ and the critical density $\rho_c$, respectively, and the red dashed line denotes the theoretical result.}
	\label{fig:fig3}
\end{figure}

To intuitively show the bifurcation, we plot the deterministic solution $S_\mathrm{det}$ of Eq.~\eqref{eq:determ-dde} and the steady-state probability distribution $P$ obtained from simulations, both in the $(\psi(t-\tau),\psi(t))$ phase plane for different $\rho_N$.
Take $\tau=0.05$ as an example.
For $\rho_s<\rho_N<\rho_c$ [Fig.~\ref{fig:fig3}(a,b)], $S_\mathrm{det}$ remains a point and $P$ remains unimodal. As density increases, $P$ broadens, consistent with stronger quasicycles and reduced polarization.
For $\rho_N>\rho_c$ [Fig.~\ref{fig:fig3}(c)], $S_\mathrm{det}$ becomes a periodic orbit and $P$ develops a ring-shaped structure that expands with density, further reducing polarization.
By contrast, for sufficiently large $\rho_N$ in the delay-free case, $S_\mathrm{det}$ remains a point and $P$ remains unimodal [Fig.~\ref{fig:fig3}(d)], which yields a narrow unimodal $\psi$ distribution and a large $\phi_N$, consistent with the normal DPR.
Furthermore, the time autocorrelation of $\psi$ in the delayed CAP system displays three distinct behaviors: overdamped relaxation in the gas, and heavily and weakly damped oscillations in the droplet periphery and core, respectively
(see SM for details).

\paragraph*{Scaled curves.--}
For quantitative comparison with simulations, we compute the theoretical polarization $\phi_\mathrm{dist}(\beta):=\langle\cos\psi\rangle$ from the stationary distribution obtained numerically from Eq.~\eqref{eq:SStateApprox2}. Combining $\phi_\mathrm{dist}(\beta)$ with the corresponding density $\rho:=\beta/[K\phi_\mathrm{dist}(\beta)]$ for each $\beta$ yields the theoretical DPR $\phi_\mathrm{dist}$--$\rho$. The theoretical curve $\phi_\mathrm{dist}$ agrees well with the simulation data $\langle\phi_N\rangle$ for $\tau=0.05$ [Fig.~\ref{fig:fig3}(e)].
Moreover, $\rho_s=\beta_s/[K\phi_\mathrm{dist}(\beta_s)]$ coincides with the density at which $\langle\phi_N\rangle$ reaches its maximum, marking the onset of the polarization decline.
Meanwhile, $\rho_c=\beta_c/[K\phi_\mathrm{dist}(\beta_c)]$ exactly coincides with the density at the most negative value of $\mathrm{d}\langle\phi_N\rangle/\mathrm{d}\rho_N$ (see Fig. S5 in SM), indicating the most rapid decrease of $\phi_N$ with $\rho_N$ near $\rho_c$.
In addition, the $\langle\phi_N\rangle$--$\rho_N$ curves for different $\tau$ (see Fig. S6 in SM) are qualitatively consistent with the $\langle\phi_L\rangle$--$\rho_L$ curves in Fig.~\ref{fig:fig1}(g).

When plotted in scaled coordinates ($\tilde{\rho}_N = (\rho_N-\rho_s)/\rho_s$, $\tilde{\phi}_N = \langle\phi_N\rangle/\langle\phi_N\rangle_{\max}$), the $\langle\phi_N\rangle$--$\rho_N$ curves for different $\tau$ merge onto a single master curve, as shown in Fig.~\ref{fig:fig3}(f). The red dashed line represents the scaled theoretical curve $\tilde\phi_\mathrm{dist}$--$\tilde\rho_N$ (see Fig. S7 in SM), which also merges onto the same master curve.
In scaled coordinates, the values of $\rho_c$ for different $\tau$ nearly coincide, implying that $\rho_s/\rho_c$ is independent of $\tau$. 
The scaling parameters reflect the degree of the DPR anomaly ($\Delta\phi$). As $\tau$ increases, the vertical scale $\langle\phi_N\rangle_{\max}$ decreases linearly, whereas the horizontal scale $\rho_s$ decays as shifted $\tau^{-1}$, as shown in the inset of Fig.~\ref{fig:fig3}(f). 
For small $\tau$, $\rho_s$ decreases much more rapidly than $\langle\phi_N\rangle_{\max}$, resulting in a much stronger horizontal compression of the master curve than vertical.
Consequently, $\langle\phi_L\rangle(\rho_{L,\max})$ in Fig.~\ref{fig:fig1}(h) drops faster than $\langle\phi_L\rangle_{\max}$ with increasing $\tau$, causing a sharp increase in $\Delta\phi$. 
For large $\tau$, the decay of $\rho_s$ slows down, resulting in a comparable vertical compression to horizontal compression. 
Hence, $\langle\phi_L\rangle(\rho_{L,\max})$ and $\langle\phi_L\rangle_{\max}$ decrease at similar rates, thereby saturating the growth of $\Delta\phi$.

\begin{figure}[!t]
	\centering
	\includegraphics[width=1\linewidth]{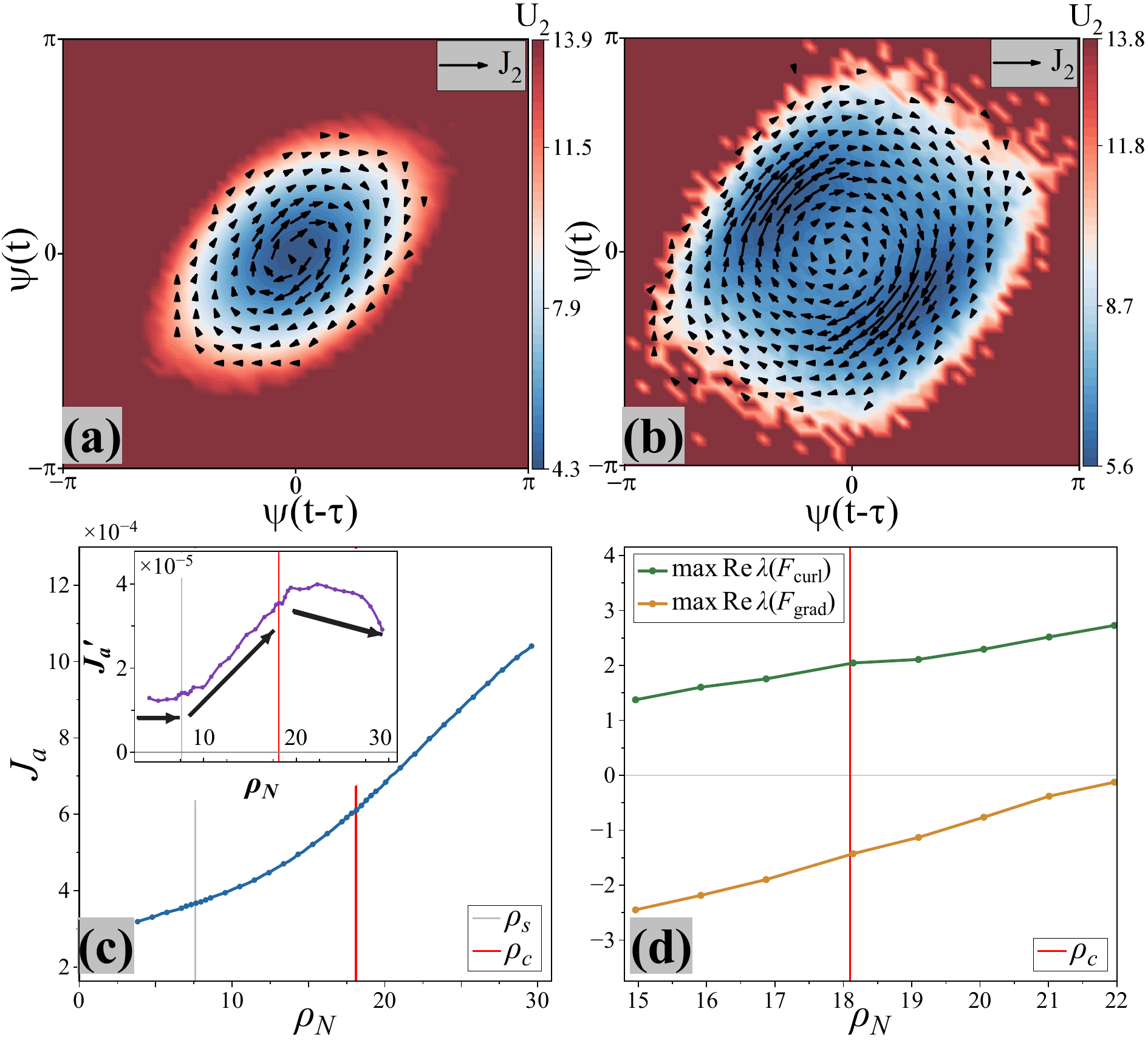}
	\caption{Landscape-flux analysis for $\tau=0.05$. (a,b) 2D potential $U_2$ and flux $\mathbf J_2$ at $\rho_N=9$ and $30$, respectively. (c) The averaged flux magnitude $J_a$; inset: $J_a'=\mathrm{d}J_a/\mathrm{d}\rho_N$. (d) $\max\operatorname{Re}\lambda(\mathbf{F}_{\mathrm{curl}})$ and $\max\operatorname{Re}\lambda(\mathbf{F}_{\mathrm{grad}})$ for the curl (green) and gradient (orange) forces at different $\rho_N$, respectively. Red and gray lines mark $\rho_c$ and $\rho_s$, respectively.}
	\label{fig:fig4}
\end{figure}

\paragraph*{Dynamical mechanism of anomalous DPR.--}
To uncover the dynamical origin of the anomalous DPR, we apply the nonequilibrium landscape-flux theory~\cite{WangPotential2008,WangLandscape2015,FangNonequilibrium2019,SuDynamical2024,YanMechanism2025,JianUncovering2025} in the $(n_\tau+1)$-dimensional delay-embedded space spanned by $\psi(t),\psi(t-\Delta t),\ldots,\psi(t-n_\tau\Delta t)$ at different $\rho_N$, where $n_\tau=\tau/\Delta t$ is the number of delay steps in the discretized system (see SM for details).
In steady state, the probability density $P$ defines the nonequilibrium potential landscape $U:=-\ln P$, which generates the gradient force $\mathbf{F}_{\mathrm{grad}}$, tending to drive the system toward potential minima.
The nonequilibrium flux $\mathbf{J}:=P\mathbf{v}$ determines the curl force $\mathbf{F}_{\mathrm{curl}}$, which drives circulation in this delay-embedded space. Here, $\mathbf{v}$ is the probability velocity field.
For visualization, we perform dimensionality reduction and construct  the two-dimensional potential $U_2=-\ln P_2$ and flux $\mathbf J_2$ in the $(\psi(t-\tau),\psi(t))$ plane (see SM for details).
Below $\rho_c$ [Fig.~\ref{fig:fig4}(a)], $\mathbf{J}_2$ circulates around the stable fixed point in the single basin of $U_2$, tending to broaden the basin and counteracting the attraction toward the fixed point.
Above $\rho_c$ [Fig.~\ref{fig:fig4}(b)], $U_2$ develops a ring-shaped valley that attracts the system toward the periodic orbit. $\mathbf{J}_2$ circulates along this valley, sustaining the oscillations.

The averaged flux magnitude, $J_a:=\langle |\mathbf{J}|\rangle$, increases with $\rho_N$ [Fig.~\ref{fig:fig4}(c)], indicating stronger circulation that contributes to the polarization decline above $\rho_s$. 
Moreover, the slope of the average flux, $J_a':=\mathrm{d}J_a/\mathrm{d}\rho_N$, reveals three distinct regimes [inset of Fig.~\ref{fig:fig4}(c)].
Below $\rho_s$, the attractive basin dominates, while both $J_a'$ and $J_a$ remain small, corresponding to the normal DPR regime.
For $\rho_s < \rho_N < \rho_c$, both $J_a'$ and $J_a$ rise sharply with $\rho_N$, so that the flux dominates, leading to a decrease in $\phi_N$ and thus to the anomalous DPR regime.
$J_a'$ peaks near $\rho_c$ and then decreases slowly, indicating that $J_a$ grows most rapidly near the bifurcation and making $J_a'$ a dynamical signature of it.
Furthermore, we evaluate $\max\operatorname{Re}\lambda$, the largest real part of the Jacobian eigenvalues for the gradient and curl forces (see SM for details).
Near $\rho_c$, $\max\operatorname{Re}\lambda(\mathbf{F}_{\mathrm{grad}})<0$ and $\max\operatorname{Re}\lambda(\mathbf{F}_{\mathrm{curl}})>0$ [Fig.~\ref{fig:fig4}(d)], revealing a competition between the stabilizing tendency of the potential landscape and the destabilizing tendency of the flux.
Together, these results support the flux as the dynamical origin of the anomalous DPR.

\begin{figure}[!t]
	\centering
	\includegraphics[width=1\linewidth]{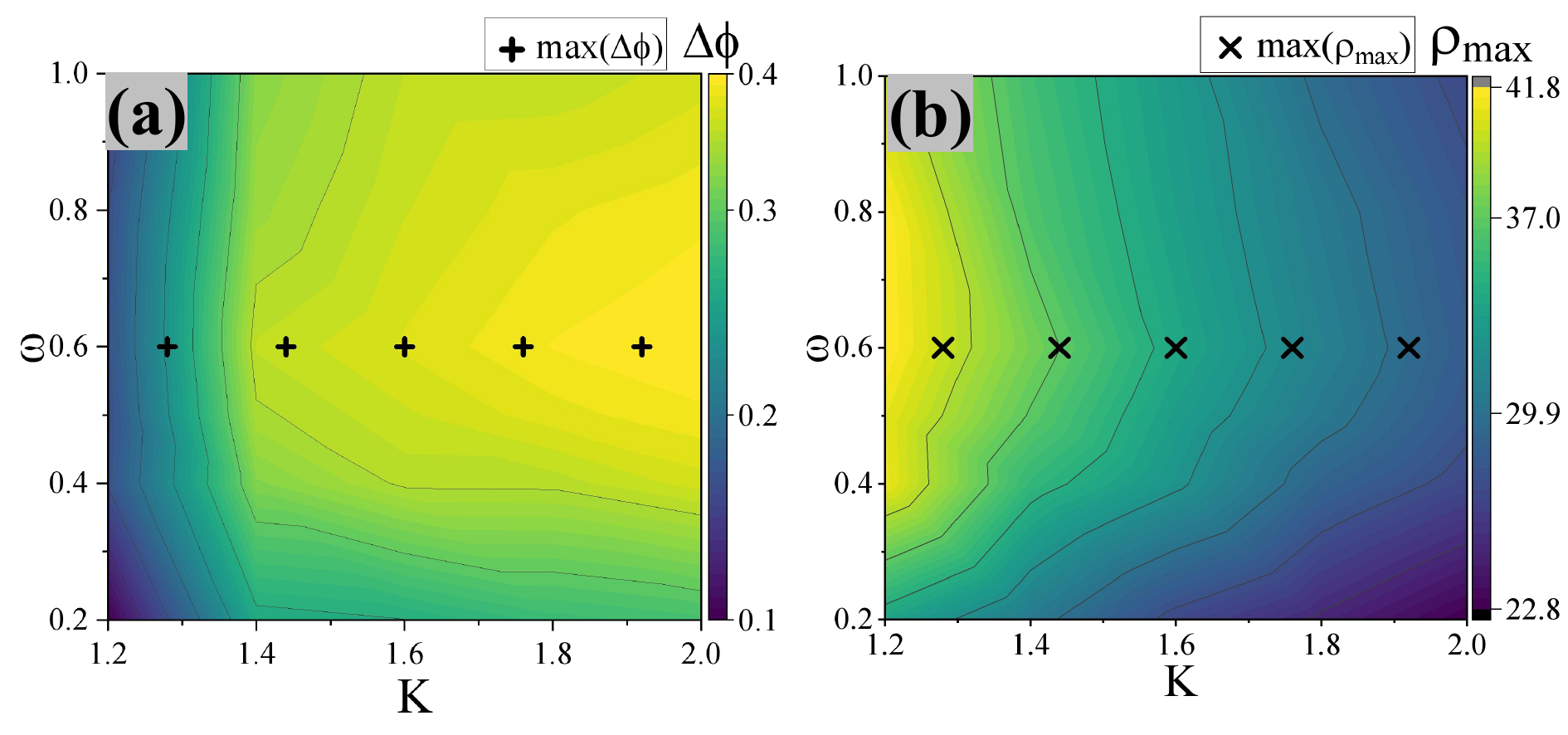}
	\caption{Influence of $K$ and $\omega$ on the anomalous DPR. (a) $\Delta\phi$ in the $K$--$\omega$ parameter space. (b) $\rho_{\max}$ in the $K$--$\omega$ parameter space. Each data point is averaged over 20 ensembles. Parameters: $\tau=0.05$, $v=1$, $\eta=1.5$.}
	\label{fig:fig5}
\end{figure}

\paragraph*{Influence of $K$ and $\omega$ on anomalous polarization.--}
In addition to $\tau$, we also examine the influence of $K$ and $\omega$ on the anomalous DPR.
Fig.~\ref{fig:fig5}(a) shows $\Delta\phi$ for different values of $K$ and $\omega$.
For fixed $\omega$, $\Delta\phi$ increases with $K$, making the DPR more anomalous. The reason is that a larger $K$ raises $\beta$, which facilitates the transition into the bifurcation regime.
For fixed $K$, $\Delta\phi$ first increases and then decreases with $\omega$, reaching a maximum $\Delta\phi_{\max}$ at $\omega^* \approx 0.6$ [Fig.~\ref{fig:fig5}(a)]. 
This non-monotonic dependence cannot be explained by the equivalent single-particle equation Eq.~\eqref{eq:SStateApprox2} alone, because it neglects spatial coupling. 
In fact, it originates from the non-monotonic dependence of $\rho_{\max}$ on $\omega$, since $\rho_{\max}$ also peaks at $\omega^*$ [Fig.~\ref{fig:fig5}(b)].
When $\omega$ is small, alignment interactions dominate the particle dynamics, allowing neighboring particles to rapidly phase-lock into a synchronously rotating macro-droplet. Due to the rotation-enhanced polarization effect \cite{LiebchenCollective2017}, faster rotation enables the droplet to absorb randomly directed particles more effectively, thereby increasing its density. Consequently, for $\omega < \omega^*$, $\rho_{\max}$ increases with $\omega$ [Fig.~\ref{fig:fig5}(b)].
However, as $\omega$ increases further, the rotation timescale becomes comparable to or even shorter than the alignment timescale \cite{LiebchenCollective2017}. This generates significant phase differences between neighboring particles, frustrating phase-locking and driving the system into the micro-droplet phase \cite{LiebchenCollective2017}. These phase differences weaken polar alignment, reduce the droplet size, and consequently decrease the droplet density. Hence, for $\omega > \omega^*$, $\rho_{\max}$ declines with $\omega$ [Fig.~\ref{fig:fig5}(b)].
Overall, $\rho_{\max}$ peaks at $\omega^*$ for fixed $K$, causing $\langle\phi_L\rangle(\rho_{\max})$ to reach its minimum, which in turn gives rise to a non-monotonic dependence of $\Delta\phi$ on $\omega$ with a maximum at the same point.

\paragraph*{Conclusion.--}
In summary, by incorporating time delay into the chiral active particle (CAP) model, we have uncovered a counterintuitive anomalous density-polarization relation (DPR) in ordered droplets: the densest core is less polarized than the lower-density periphery. 
As density increases beyond the optimal density $\rho_s$, noise-sustained quasicycles around the stable fixed point broaden the deviation-angle distribution and reduce polarization. At the critical density $\rho_c$, a supercritical discrete Hopf bifurcation gives rise to a periodic orbit, whose self-sustained oscillations further reduce polarization.
Quantitatively, the density--polarization curve obtained from theory agrees well with those obtained from simulations. 
Furthermore, $\rho_s$ and $\rho_c$ decrease with increasing delay, bringing the bifurcation to lower densities, which makes the anomalous DPR more pronounced, in line with the simulation observation.
More importantly, the nonequilibrium landscape-flux analysis identifies the nonequilibrium flux as the dynamical origin of the anomalous DPR and the density slope of the flux as a dynamical signature of the bifurcation.
In addition, we explore the effects of alignment strength and chirality strength on the anomalous DPR.
Our findings establish time delay as a mechanism for anomalous DPR and advance the theoretical understanding of collective behavior in chiral active matter. The resulting predictions are experimentally testable using programmable light-controlled robots \cite{MijalkovEngineering2016,LeymanTuning2018} and may guide the design of collective patterns.
More broadly, our theoretical framework offers a route toward a unified understanding of density-modulated collective dynamics across diverse nonequilibrium systems, including polar active liquids \cite{GeyerFreezing2019}, chemotactic bacteria \cite{ColinChemotactic2019}, pedestrian crowds \cite{MaSpontaneous2021,MaUnraveling2025}, and confluent epithelia \cite{TangCollective2026}.

Dongrun Jian, Jie Su, and Jun Wang gratefully acknowledge funding from National Natural Science Foundation of China (Nos. 12404237 and 12234019), the Zhejiang Key Laboratory of Soft Matter Biomedical Materials (Nos. 2025ZY01036 and 2025E10072) and Wenzhou Institute of Biomaterials and Engineering (No. WIUCASQD2022012).

\bibliography{DCAP}%

\end{document}


\maketitle

\noindent
$^{1}$Department of Physics, Nanjing University, Nanjing, Jiangsu 210093, China\\
$^{2}$Center for Theoretical Interdisciplinary Sciences, Wenzhou Institute, University of Chinese Academy of Sciences, Wenzhou 325001, China\\
$^{3}$Zhejiang Key Laboratory of Soft Matter Biomedical Materials, Wenzhou Institute, University of Chinese Academy of Sciences, Wenzhou, Zhejiang 325000, China\\
$^{4}$Department of Chemistry and of Physics and Astronomy, State University of New York of Stony Brook, Stony Brook, New York 11794, USA

\vspace{12pt}

\section{Derivation of the dynamical equation for the deviation angle}
\subsection{The parameter transformation}
For a given particle $i$, let $N_i$ denote the set of all particles within the alignment radius $R$ of particle $i$, including particle $i$ itself. 
We define the average orientation vector of the particles in $N_i$ as:
\begin{equation}
\label{eq:phi-alpha}
\phi_{N}^i e^{\mathrm{i}\alpha_{i}}=\frac{1}{|N_i|}\sum_{j\in N_i}e^{\mathrm{i}\theta_j},
\end{equation}
where $\theta_j$ denotes the orientation angle of particle $j$, $\phi_{N}^i$ is the neighborhood polarization, $|N_i|$ is the number of particles in set $N_i$, and $\alpha_{i}$ is the average angle over $N_i$. Multiplying both sides by $\frac{|N_i|}{\pi R^2}e^{-\mathrm{i}\theta_i}$, taking the imaginary part, and replacing $\frac{|N_i|}{\pi R^2}$ by the neighborhood density $\rho_{N}^i$, we obtain the parameter transformation:
\begin{equation}
	\label{eq:para-trans}
	\rho_{N}^i\phi_{N}^i\sin(\alpha_{i}-\theta_i)=\frac{1}{\pi R^2}\sum_{j\in N_i}\sin(\theta_j-\theta_i),
\end{equation}
where $\theta_i-\alpha_i$ corresponds to the deviation angle $\psi_i$ used in the main text.

\subsection{The steady-state approximation}

\begin{figure}[htbp]
	\centering
	\includegraphics[width=0.92\linewidth]{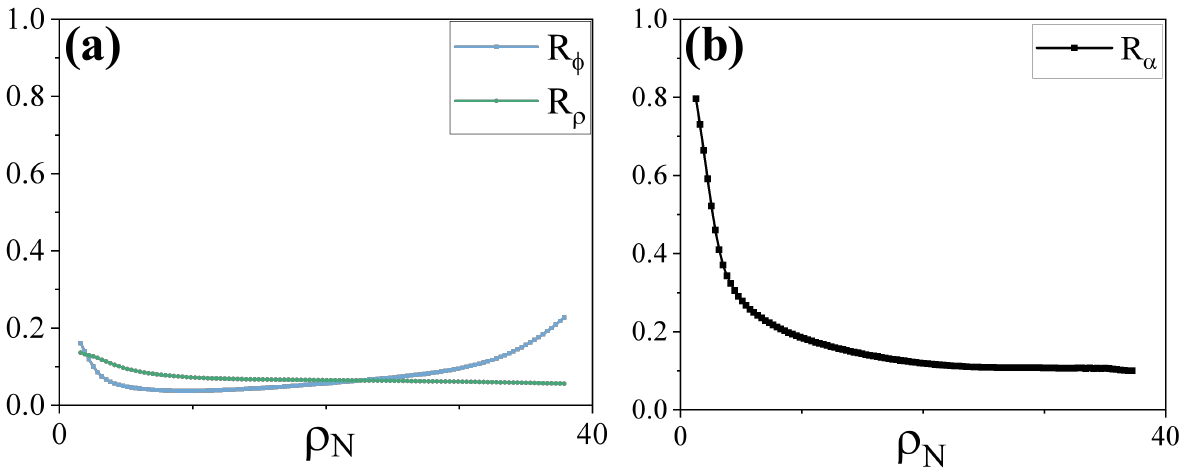}
	\caption{
		(a) Relative variations of the local polarization $R_\phi$ and the local density $R_\rho$ as functions of $\rho_N$. 
		(b) Ratio of $|\dot\alpha-\omega|$ to $|\dot\psi|$, $R_\alpha$, as a function of $\rho_N$.
	}
	\label{fig:fig-Approxi}
\end{figure}

Inside the steady-state droplet ($\rho_N^i>\rho_0=5$), the neighborhood density $\rho_N^i(t)$ and polarization $\phi_N^i(t)$ vary little over timescales of order $\tau$. We quantify their relative variations by
\begin{equation}
R_\phi:=\frac{\langle|\delta\phi_N^i(t)|\rangle}{\langle|\phi_N^i(t)|\rangle},
\qquad
R_\rho:=\frac{\langle|\delta\rho_N^i(t)|\rangle}{\langle|\rho_N^i(t)|\rangle},
\end{equation}
where $\delta\phi_N^i(t):=\phi_N^i(t+\tau)-\phi_N^i(t)$, $\delta\rho_N^i(t):=\rho_N^i(t+\tau)-\rho_N^i(t)$,
and the averages are taken over time and particles. As shown in Fig.~\ref{fig:fig-Approxi}(a), both $R_\phi$ and $R_\rho$ are much smaller than 1 for $\rho_N>\rho_0$, indicating that $\phi_N^i$ and $\rho_N^i$ remain approximately constant over time intervals of order $\tau$. We therefore approximate them as time-independent, $\rho_N^i(t)=\rho_N^i$, $\phi_N^i(t)=\phi_N^i$.

Defining the deviation angle $\psi_i(t):=\theta_i(t)-\alpha_i(t)$, the angular equation Eq.~(1) of the main text becomes
\begin{equation}
\label{SStateApprox1}
\dot{\psi}_i(t)= -\beta_i\sin(\psi_i(t-\tau))+\sqrt{2\eta}\zeta_i(t)-\bigl[\dot{\alpha}_i(t)-\omega\bigr],
\end{equation}
where $\beta_i:=K\rho_N^i\phi_N^i$ is the effective alignment strength of the neighborhood. Inside the droplet, $\dot{\alpha}_i(t)$ fluctuates weakly around $\omega$. To quantify the magnitude of this fluctuation relative to $\dot{\psi}_i(t)$, we define
\begin{equation}
R_\alpha:=\frac{\langle|\dot{\alpha}_i(t)-\omega|\rangle}{\langle|\dot{\psi}_i(t)|\rangle},
\end{equation}
where the averages are taken over time and particles.
As shown in Fig.~\ref{fig:fig-Approxi}(b), $R_\alpha\ll1$ for $\rho_N>\rho_0$, indicating that $|\dot{\alpha}_i(t)-\omega|\ll|\dot{\psi}_i(t)|$. We may therefore neglect the term $\dot{\alpha}_i(t)-\omega$, yielding
\begin{equation}
\label{eq:SStateApprox2}
\dot{\psi}_i(t)= -\beta_i\sin(\psi_i(t-\tau))+\sqrt{2\eta}\zeta_i(t).
\end{equation}

\section{Bifurcation analysis of the deterministic delayed difference equation}
To reveal the dynamical mechanism of anomalous DPR, we consider the deterministic part of Eq.~\eqref{eq:SStateApprox2}. We define $\psi_k:=\psi(k\Delta t)$ and take $\tau=n_\tau\Delta t$, where the integer $n_\tau\geq0$ is the number of delay steps. Discretizing in time yields a single-variable delay-difference equation:
\begin{equation}
\label{eq:determ-dde}
\psi_{k+1}-\psi_k=-\beta\Delta t\sin\psi_{k-n_\tau}.
\end{equation}
The fixed points of Eq.~\eqref{eq:determ-dde} satisfy $\sin\psi^*=0$.
We restrict the deviation angle $\psi$ to the range $\psi\in(-\pi,\pi]$, hence the fixed points are $\psi^*=0$ or $\pi$.

Introducing a small perturbation $u_k:=u(k\Delta t)$, let $\psi_k=\psi^*+u_k$. The fixed point $\psi^*=\pi$ is unstable for any $\beta>0$.
Linearizing Eq.~\eqref{eq:determ-dde} around $\psi^*=0$ yields the linear difference equation:
\begin{equation}
	\label{eq:linear-dde}
	u_{k+1}-u_k=-\beta\Delta t\,u_{k-n_\tau}.
\end{equation}

Substituting the ansatz $u_k=e^{\lambda k\Delta t}$, where $\lambda$ is the growth rate per unit time, yields the characteristic equation:
\begin{equation}
	\label{eq:element-eq}
e^{\lambda(n_\tau+1)\Delta t}-e^{\lambda n_\tau\Delta t}+\beta\Delta t=0.
\end{equation}
We define the characteristic function by
\begin{equation}
\label{eq:characteristic-function}
\mathcal L(\lambda):=\frac{e^{\lambda\Delta t}-1}{\Delta t}+\beta e^{-\lambda\tau}.
\end{equation}
Equation~\eqref{eq:element-eq} can then be written as $\mathcal L(\lambda)=0$.

Taking the critical oscillatory eigenvalue $\lambda=i\Omega$, where $\Omega$ denotes the oscillation frequency, and substituting into Eq.~\eqref{eq:element-eq}, the phase and amplitude conditions give:
\begin{equation}
	\begin{aligned}
		\Omega\bigl(n_\tau+\tfrac{1}{2}\bigr)\Delta t&=\pm\frac{\pi}{2}\pmod{2\pi},\\
		\beta\Delta t &=2\left|\sin\!\left(\frac{\Omega\Delta t}{2}\right)\right|,
	\end{aligned}
\end{equation}
where the $\pm$ sign is matched with the sign of $\sin(\Omega\Delta t/2)$ (plus for $\sin(\Omega\Delta t/2)>0$, minus otherwise), so that $e^{i\Omega n_\tau\Delta t}\bigl(e^{i\Omega\Delta t}-1\bigr)=-\beta\Delta t$ is a negative real number.
Taking the lowest-frequency branch $\Omega_c\Delta t=\frac{\pi}{2n_\tau+1}$ (with the plus sign), we obtain the bifurcation critical point:
\begin{equation}
	\label{eq:bc}
	\beta_c =\frac{1}{\Delta t}2\sin\!\left(\frac{\pi}{2(2n_\tau+1)}\right).
\end{equation}
The following discrete Hopf analysis concerns $n_\tau>0$.
Let $\lambda=\lambda(\beta)$. Differentiating the characteristic equation~\eqref{eq:element-eq} with respect to $\beta$ yields
\begin{equation}
\label{eq:root-derivative}
\begin{aligned}
\frac{\mathrm{d}\lambda}{\mathrm{d}\beta}
&=-\frac{e^{-\lambda\tau}}{e^{\lambda\Delta t}-\beta\tau e^{-\lambda\tau}}\\
&=-\frac{1}{(n_\tau+1)e^{\lambda(n_\tau+1)\Delta t}-n_\tau e^{\lambda n_\tau\Delta t}}.
\end{aligned}
\end{equation}
Evaluating at the critical eigenvalue $\lambda=i\frac{\pi}{(2n_\tau+1)\Delta t}$, we obtain $\mathrm{Re}(\mathrm{d}\lambda/\mathrm{d}\beta)>0$, indicating that at the critical point $\beta_c$ the eigenvalue crosses the imaginary axis, confirming that this bifurcation is a discrete Hopf bifurcation.

Since the solution near the bifurcation point has its principal frequency near $\Omega_c$, we project the solution onto the critical eigenmode $e^{i\Omega_c k\Delta t}$ (the complex amplitude representation):
\begin{equation}
\label{eq:critical-mode-expansion}
\psi_k = z_k e^{i\Omega_c k\Delta t} + \bar{z}_k e^{-i\Omega_c k\Delta t} + O(|z_k|^2).
\end{equation}
Reducing the nonlinear delay-difference equation~\eqref{eq:determ-dde} to the two-dimensional center manifold and retaining terms through cubic order, we obtain the normal form equation:
\begin{equation}
		z_{k+1}=z_k+\kappa(\beta-\beta_c)z_k+\nu z_k|z_k|^2+\cdots,
\end{equation}
where $z_k$ is the complex amplitude at step $k$, and $\kappa$ and $\nu$ denote the linear growth coefficient and the cubic saturation coefficient, respectively. Their real parts are explicitly given by
$\mathrm{Re}\,\kappa=\frac{(2n_\tau+1)\beta_c\Delta t^2}{2d}>0$ and $\mathrm{Re}\,\nu=-\frac{(2n_\tau+1)\beta_c^2\Delta t^2}{4d}<0,$
with $d:=\frac{(2n_\tau+1)^2\beta_c^2\Delta t^2}{4}+\cos^2\frac{\Omega_c\Delta t}{2}$. The first Lyapunov coefficient is $l_1:=\mathrm{Re}\,\nu$, whose negativity indicates that the system undergoes a supercritical discrete Hopf bifurcation at the critical point. Since the principal modal frequency is a rational fraction of $\pi$, resonance occurs, forming a periodic orbit in delay-embedded space with a period of $2(2n_\tau+1)\Delta t$ (which can be regarded as a discrete version of a limit cycle).

In summary, the dynamical properties of the deterministic delayed difference equation are:
\begin{enumerate}
	\item For $0<\beta<\beta_c$, the fixed point $\psi^*=0$ is locally asymptotically stable.
	\item At $\beta=\beta_c$, the system undergoes a supercritical discrete Hopf bifurcation.
	\item For $\beta>\beta_c$, the fixed point $\psi^*=0$ loses stability and a period-$2(2n_\tau+1)$ orbit emerges, which is the ``limit cycle'' of the discrete system.
\end{enumerate}

\section{Steady-state variance and the maximum point of polarization}

As in the main text, we define the theoretical polarization as $\phi_{\mathrm{dist}}=\langle\cos\psi\rangle$, where the average is taken over the stationary deviation-angle distribution at a given neighborhood density. Throughout this section, we consider fluctuations around the stable deterministic fixed point $\psi^*=0$ for $0<\beta<\beta_c$. Within the linear approximation, this distribution is a zero-mean Gaussian~\cite{FrankFokkerPlanck2003} with variance $V:=\mathrm{Var}(\psi)$, yielding
\begin{equation}
\phi_{\mathrm{dist}}\simeq e^{-V/2}.
\label{eq:polarization-variance}
\end{equation}
Within this approximation, the polarization maximum corresponds to the variance minimum, and increasing variance implies decreasing polarization. We derive the stationary variance below and interpret its increase as $\beta$ approaches $\beta_c$ from below.

\subsection{Steady-state variance of the continuous system}
Taking the continuous limit $\Delta t\to 0$ at fixed $\tau>0$ in the discrete critical value Eq.~\eqref{eq:bc}, the bifurcation point becomes $\beta_c=\frac{\pi}{2\tau}$. Linearizing the continuous dynamical equation Eq.~\eqref{eq:SStateApprox2} around $\psi^*=0$ yields:
\begin{equation}
	\label{eq:linear-dde-continuous}
	\dot{u}(t)=-\beta u(t-\tau)+\sqrt{2\eta}\zeta(t).
\end{equation}
Its stationary variance $V(\beta)$ can be derived from the Fokker--Planck equation \cite{KuchlerLangevins1992,FrankFokkerPlanck2003}:
\begin{equation}
	V(\beta) = \frac{\eta}{\beta}\frac{1 + \sin (\beta \tau)}{\cos (\beta \tau)},
	\quad 0<\beta<\beta_c.
\end{equation}
Minimizing $V(\beta)$ with respect to $\beta$, i.e., setting $V''(\beta)>0$ and $V'(\beta)=0$, gives $\cos(\beta_s \tau) = \beta_s \tau$, whose solution is the Dottie number $\beta_s\tau=D\approx0.739085$. Thus we obtain:
\begin{equation}
	\frac{\beta_s}{\beta_c}=2D/\pi\approx0.4705.
\end{equation}
The ratio of the minimum point $\beta_s$ to the bifurcation point $\beta_c$ is a constant.

\subsection{Steady-state variance of the discrete system}

We now consider the discrete-time case. Writing $u_k:=u(k\Delta t)$, the Euler--Maruyama discretization of Eq.~\eqref{eq:linear-dde-continuous} in the stable fixed-point regime ($0<\beta<\beta_c$) reads:
\begin{equation}
	\label{eq:stoch-dde}
	u_{k+1}-u_k=-\beta\Delta t\,u_{k-n_\tau}+\sqrt{2\eta\Delta t}\,\xi_k.
\end{equation}
Here $\xi_k$ are independent standard Gaussian variables, with $\langle\xi_k\rangle=0$ and $\langle\xi_k\xi_\ell\rangle=\delta_{k\ell}$.

We now construct the Yule--Walker equations. For convenience, define $\gamma_j := \mathrm{Cov}\bigl(u_k,u_{k-j}\bigr)$, $\sigma^2 := \mathrm{Var}\bigl(\sqrt{2\eta\Delta t}\,\xi_k\bigr)=2\eta\Delta t$, and $\beta' := \beta\Delta t$. Multiplying Eq.~\eqref{eq:stoch-dde} by $u_{k+1}$ and taking the expectation yields:
\begin{equation}
	\gamma_0-\gamma_1=-\beta'\gamma_{n_\tau+1}+\sigma^2.
\end{equation}
Since $\xi_k$ is independent of $u_{k-j}$ for $j\ge 0$, multiplying Eq.~\eqref{eq:stoch-dde} by $u_{k-j}$ and taking the expectation gives:
\begin{equation}
	\gamma_{j+1}-\gamma_j=-\beta'\gamma_{j-n_\tau},\quad j=0,1,\dots,n_\tau.
\end{equation}

Combined with the symmetry $\gamma_j=\gamma_{-j}$, we obtain a closed linear system for $\gamma_0,\gamma_1,\dots,\gamma_{n_\tau+1}$. Solving for $\gamma_0$ yields the steady-state variance $V_{n_\tau}(\beta')$. $V_{n_\tau}(\beta\Delta t)$ is a rational function of $\beta$ and possesses a minimum point $\beta_s$ in the interval $0<\beta<\beta_c$. The steady-state variance minimum points $\beta_s$ for different $n_\tau$ are listed in Supplementary Table~\ref{tab:n-beta}.

\begin{table}[htbp]
	\centering
	\begin{tabular}{cccc}
		\toprule
		$n_\tau$ & $\beta_s$ & $\beta_c$ & $\beta_s/\beta_c$ \\
		\midrule
		1 & 46.557  & 100      & 0.4656  \\
		2 & 28.954  & 61.803   & 0.4685  \\
		3 & 20.892  & 44.504   & 0.4694  \\
		4 & 16.318  & 34.730   & 0.4699  \\
		5 & 13.380  & 28.463   & 0.4701  \\
		\bottomrule
	\end{tabular}
	\caption{Steady-state variance minimum point $\beta_s$ and its comparison with the critical value $\beta_c$ for different delay steps $n_\tau$. All values are computed with $\Delta t=0.01$.}
	\label{tab:n-beta}
\end{table}

As $n_\tau$ increases, $\beta_s/\beta_c$ approaches the constant value of the continuous limit ($\approx 0.4705$). Moreover, in the continuous limit $\Delta t\to 0$ with $\tau=n_\tau\Delta t$ fixed, the Yule--Walker solution reproduces the continuous variance of the previous subsection, $V_{n_\tau}(\beta\Delta t)\to V(\beta)$, and correspondingly $\beta_s/\beta_c\to 2D/\pi$.
The steady-state variance functions $V_{n_\tau}(\beta')$ for discrete system are as follows:

For $n_\tau=1$:
\begin{equation}
	V_{1}(\beta')=\sigma^2 \frac{1+\beta'}{\beta'(2-\beta'-\beta'^{2})}.
\end{equation}

For $n_\tau=2$:
\begin{equation}
	V_{2}(\beta')=\sigma^2 \frac{1+\beta'-\beta'^{2}}{\beta'(2 - 3 \beta' - \beta'^{2} + \beta'^{3})}.
\end{equation}

For $n_\tau=3$:
\begin{equation}
	V_{3}(\beta')=\sigma^{2} \frac{1+2\beta'-\beta'^{2}-\beta'^{3}}{\beta' (2-3 \beta'-4 \beta'^{2}+\beta'^{3}+\beta'^{4})}.
\end{equation}

For $n_\tau=4$:
\begin{equation}
	V_{4}(\beta')=\sigma^{2} \frac{1+2\beta'-3 \beta'^{2}-\beta'^{3}+\beta'^{4}}{\beta'(2 -5 \beta'-4 \beta'^{2}+5 \beta'^{3}+\beta'^{4}-\beta'^{5})}.
\end{equation}

For $n_\tau=5$:
\begin{equation}
	V_{5} (\beta') = \sigma^{2} \frac{1 + 3 \beta' - 3 \beta'^{2} - 4 \beta'^{3} + \beta'^{4} + \beta'^{5}}{\beta' (2 - 5 \beta' - 9 \beta'^{2} + 5 \beta'^{3} + 6 \beta'^{4} - \beta'^{5} - \beta'^{6})}.
\end{equation}

For $n_\tau=0$, the continuous-time variance $V_0=\eta/\beta$ decreases monotonically with $\beta$. The finite minimum $\beta_s=1/\Delta t$ arises from Euler--Maruyama discretization and diverges as $\Delta t\to0$. To avoid the resulting variance upturn, the delay-free comparison should satisfy $\beta\Delta t<1$.

\subsection{Noise-sustained oscillations below the bifurcation point}

\begin{figure}[htbp]
\centering
\includegraphics[width=0.78\linewidth]{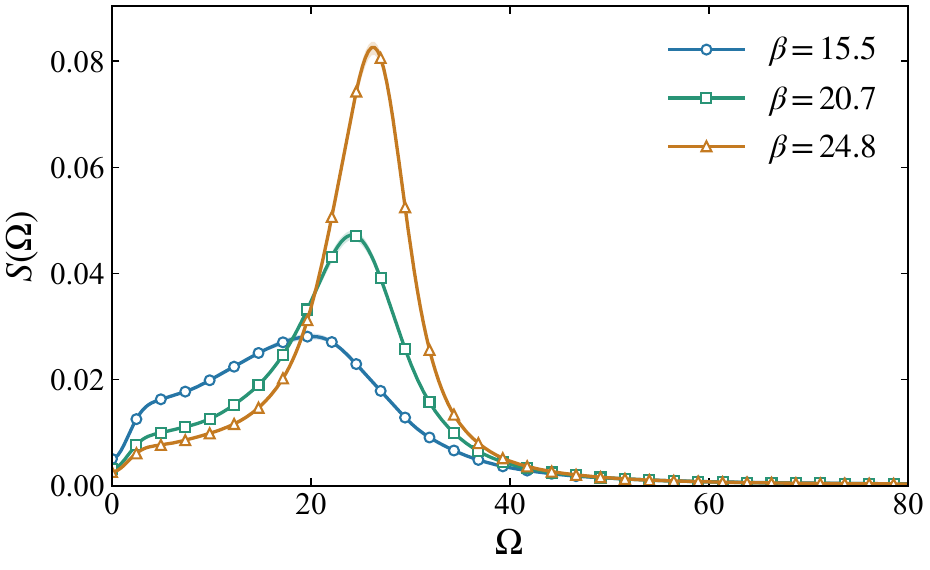}
\caption{Deviation-angle power spectra $S(\Omega)$ obtained from simulation data for $\beta_s<\beta<\beta_c$. Parameters: $\tau=0.05$, $K=2$, $\omega=0.5$, and $\eta=1.5$.}
\label{fig:power-spectrum}
\end{figure}

For $\beta_s<\beta<\beta_c$, the deterministic fixed point remains stable, but noise continuously excites the damped oscillatory mode that becomes unstable at the bifurcation. These noise-sustained oscillations are termed \textit{quasicycles}~\cite{LugoQuasicycles2008,McKanePredator2005}. To understand how these fluctuations reduce polarization, we examine their power spectrum $S(\Omega)$. Fourier transforming Eq.~\eqref{eq:stoch-dde} gives
\begin{equation}
S(\Omega)=\frac{2\eta}{|\mathcal L(i\Omega)|^2},
\label{eq:noise-spectrum-discrete}
\end{equation}
where $\mathcal L$ is the characteristic function defined in Eq.~\eqref{eq:characteristic-function}, and $|\mathcal L(i\Omega)|^{-2}$ gives the squared response to a unit harmonic perturbation at angular frequency~$\Omega$. The variance is the integrated spectral power: $V=\frac{1}{2\pi}\int_{-\pi/\Delta t}^{\pi/\Delta t}S(\Omega)\,\mathrm{d}\Omega$.

The oscillatory mode that loses stability at $\beta_c$ has a characteristic root $\lambda_m=-\Gamma_m+i\Omega_m$, satisfying $\mathcal L(\lambda_m)=0$. Here, $\Gamma_m>0$ is its decay rate, and $\Omega_m$ is its oscillation frequency, distinct from the spectral frequency variable $\Omega$. As $\beta$ approaches $\beta_c$, $\Gamma_m\to0$ and $\Omega_m\to\Omega_c$. In the linear approximation, at fixed noise intensity, the resonant peaks have heights proportional to $\Gamma_m^{-2}$ and widths proportional to $\Gamma_m$, so their integrated power grows as $\Gamma_m^{-1}$. Thus, weaker damping amplifies the response to noise, increasing the variance of $\psi$ and reducing polarization. The predicted divergence at the threshold signals the breakdown of the linear approximation.

Figure~\ref{fig:power-spectrum} shows the power spectra $S(\Omega)$ obtained from simulation data at three effective alignment strengths below $\beta_c$. As $\beta$ increases, the peak of $S(\Omega)$ becomes more pronounced and the integrated spectral power increases. These observations support that noise sustains increasingly strong oscillatory fluctuations before the deterministic fixed point loses stability, broadening the deviation-angle distribution and lowering polarization.

\section{Monotonicity of $\beta(\rho_N)$}

\begin{figure}[htbp]
	\centering
	\includegraphics[width=0.6\linewidth]{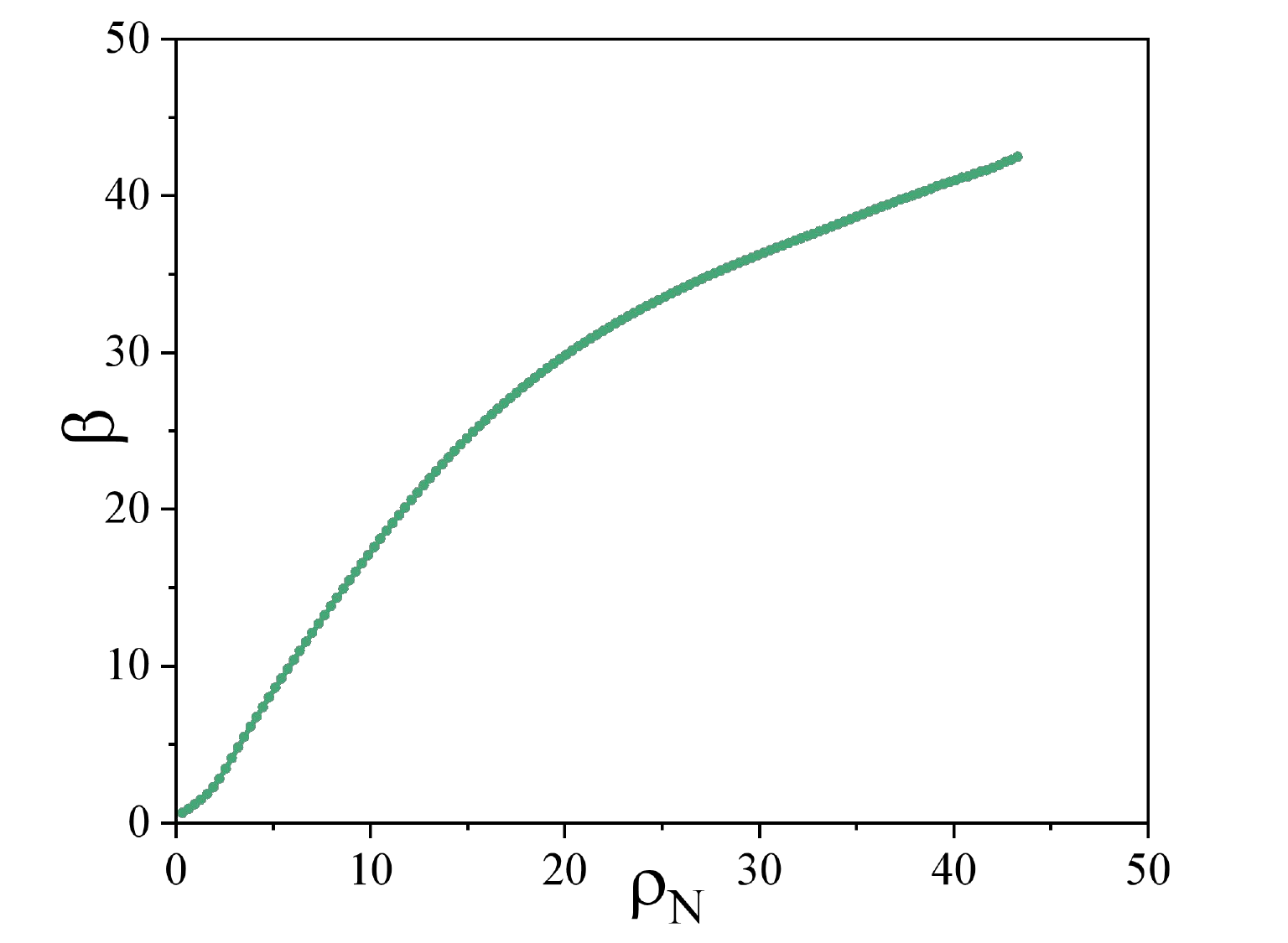}
	\caption{The equivalent alignment strength $\beta$ versus the local density $\rho_N$ for $\tau=0.05$.}
	\label{fig:fig-Beta(rho)}
\end{figure}

Taking $\tau=0.05$ as an example (Fig.~\ref{fig:fig-Beta(rho)}), the simulation data all fall within the monotonically increasing regime of $\beta(\rho_N)$, ensuring a one-to-one correspondence between $\beta$ and $\rho_N$.

\section{Correlation in the delayed chiral active particle system}

To capture the relaxation dynamics of $\psi$, we compute its time autocorrelation function:
\begin{equation}
\label{eq:psi-acf}
C_\psi(t)=\left\langle \cos\bigl(\psi_i(t_0+t)-\psi_i(t_0)\bigr)\right\rangle_{i,t_0,\rho_N\in \mathrm{W}},
\end{equation}
where the average is taken over time and over all particles $i$ whose neighborhood density $\rho_N$ lies within a prescribed window $\mathrm{W}$. 
We denote $C^g_\psi(t)$, $C^p_\psi(t)$, and $C^c_\psi(t)$ for the gas phase ($\mathrm{W}=(0,\rho_0)$), the droplet periphery ($\mathrm{W}=(\rho_0,\rho_c]$), and the droplet core ($\mathrm{W}=(\rho_c,\rho_{\max}]$), respectively.
In the stationary regime, $C_\psi(t)$ approaches a long-time plateau:
\begin{equation}
\label{eq:psi-plateau}
C_\infty=\langle\cos\psi\rangle^2,
\end{equation}
which corresponds to the theoretical average polarization $\phi_{\mathrm{dist}}$ of the particles in the window. To clearly compare different correlation behaviors, we define the normalized autocorrelation $\tilde C_\psi(t)$:
\begin{equation}
\label{eq:psi-norm}
\tilde C_\psi(t)=\frac{C_\psi(t)-C_\infty}{1-C_\infty}.
\end{equation}

\begin{figure}[htbp]
	\centering
	\includegraphics[width=1\linewidth]{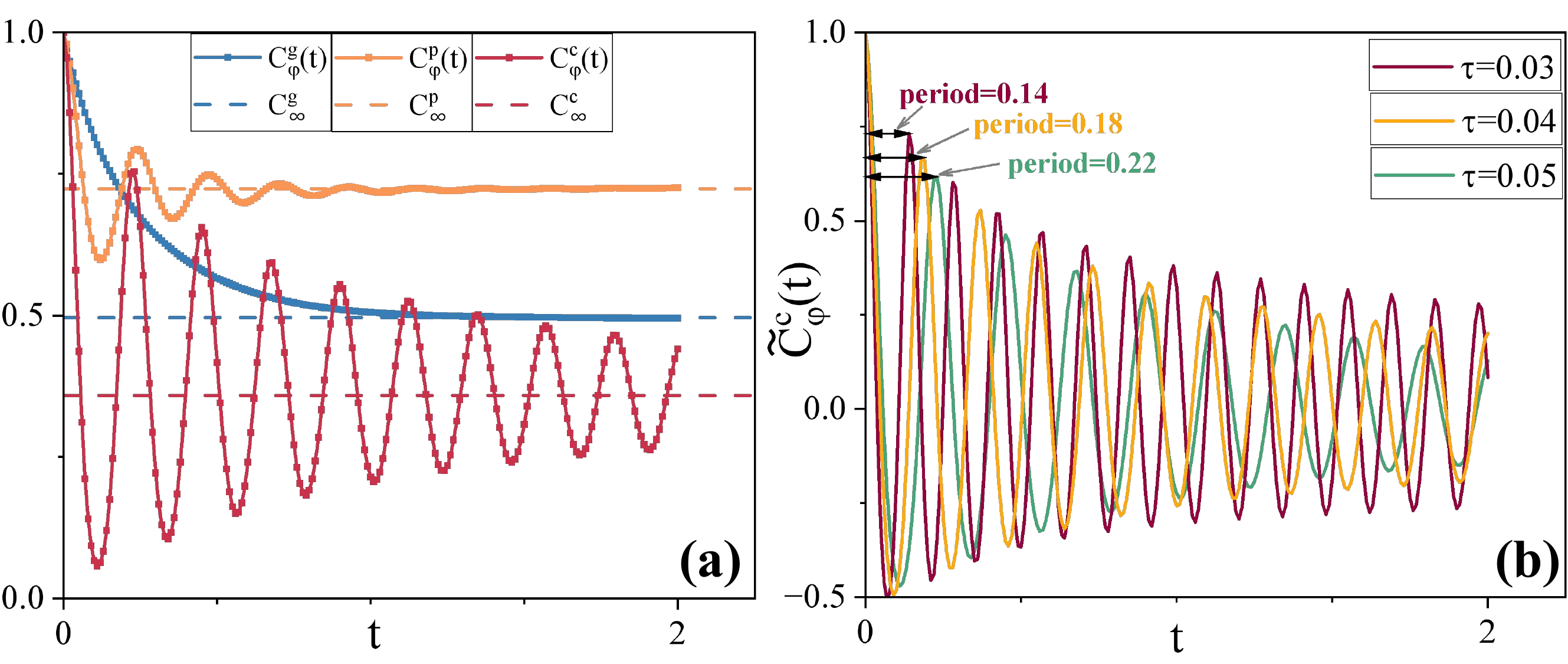}
	\caption{
		Time autocorrelations of the deviation angle $\psi$ at $\tau=0.05$.
		(a) Autocorrelation $C_\psi(t)$ (solid) with the plateau value $C_\infty$ (dashed). The superscripts $g$, $p$, and $c$ denote the gas, droplet-periphery, and droplet-core regimes, respectively.
		(b) Normalized autocorrelation $\tilde C^c_\psi(t)$ for different delays $\tau$.
	}
	\label{fig:fig-CorrelationPsi}
\end{figure}

Take $\tau=0.05$ as an example.
The autocorrelation functions $C_\psi(t)$ for three different density windows exhibit different relaxation behaviors in Fig.~\ref{fig:fig-CorrelationPsi} (a).
$C^c_\psi(t)$ shows a weakly damped oscillation that does not vanish within the observation time. This oscillation with period 0.22 is the temporal fingerprint of the period-$22$ orbit created by the bifurcation in the droplet core. 
$C^p_\psi(t)$ exhibits a heavily damped oscillation: its period is the same as that of $C^c_\psi(t)$, but the envelope shrinks quickly toward $C^p_\infty$, corresponding to the complex eigenvalues $\lambda$ with $\mathrm{Re}\,\lambda<0$ of the linearized dynamical equation Eq.~\eqref{eq:linear-dde} (the stable fixed-point regime) in the droplet periphery.
For the gas phase, $C^g_\psi(t)$ exhibits an overdamped relaxation, indicating that $\psi$ decorrelates from its history on a short timescale.
Although Eq.~\eqref{eq:linear-dde} is only valid for the droplet phase ($\rho_N>\rho_0$), it can still qualitatively explain that the dominant eigenvalue $\lambda$ becomes real and negative at low density, leading to the overdamped relaxation toward $C^g_\infty$. Moreover, as the density increases, the plateau $C_\infty$ first increases and then decreases, signaling the anomalous density--polarization relation reported in the main text.

In Fig.~\ref{fig:fig-CorrelationPsi} (b), we present the normalized autocorrelations in the droplet core $\tilde C^c_\psi(t)$ for different delays $\tau$. The oscillation period $T_c$ can be written as $2(2n_\tau+1)\Delta t$, precisely consistent with the period-$2(2n_\tau+1)$ orbit. Moreover, the oscillation envelope decays faster for larger $\tau$, indicating that the sustained oscillation is damped more strongly as the delay increases.

\section{Derivative $\mathrm{d}\langle\phi_N\rangle/\mathrm{d}\rho_N$ versus the density $\rho_N$}

\begin{figure}[htbp]
	\centering
	\includegraphics[width=0.6\linewidth]{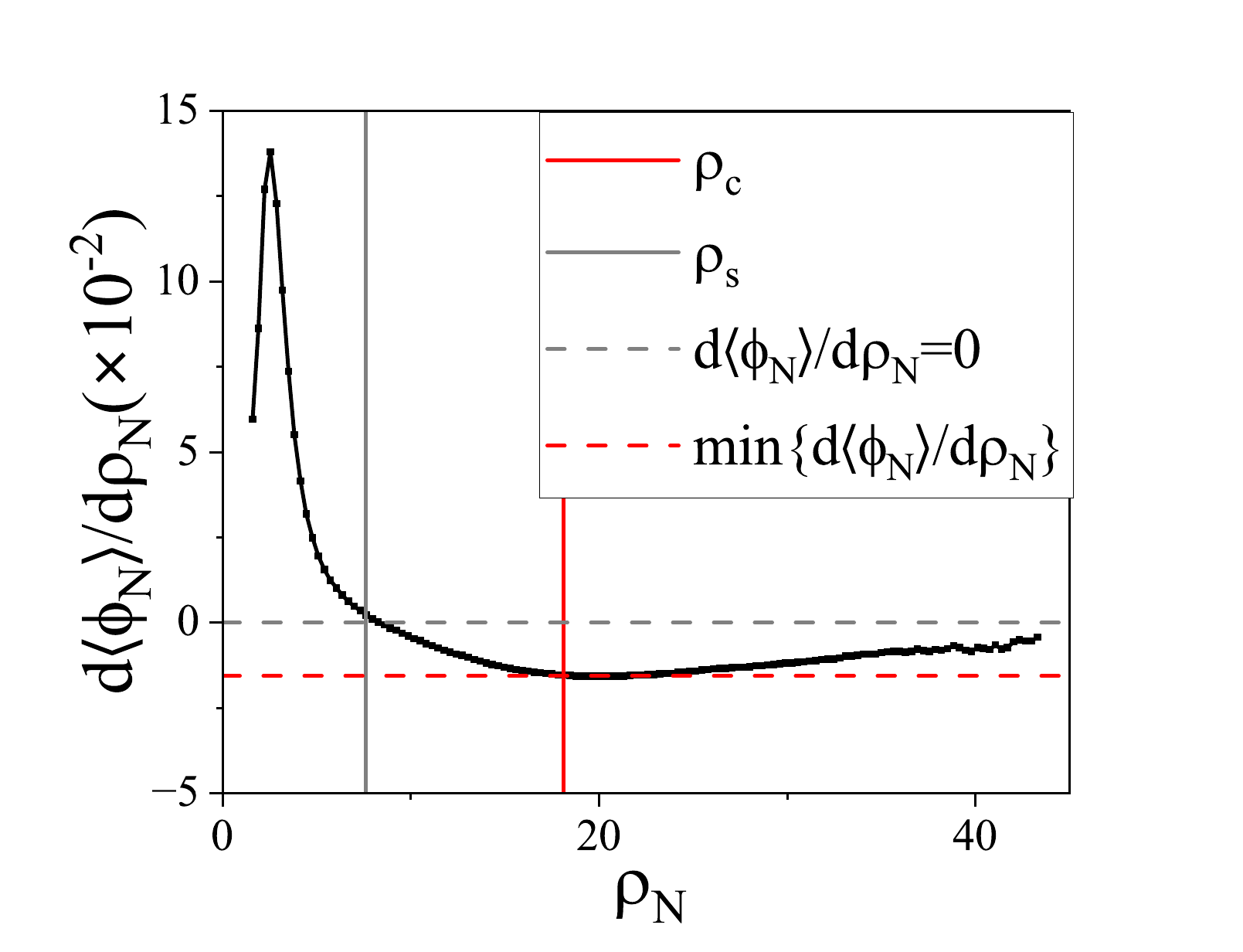}
	\caption{Derivative of the average polarization, $\mathrm{d}\langle\phi_N\rangle/\mathrm{d}\rho_N$, as a function of the density $\rho_N$ for $\tau=0.05$.}
	\label{fig:fig-Derivative}
\end{figure}

In Fig.~\ref{fig:fig-Derivative}, the theoretical polarization-maximum density $\rho_s=7.6$ lies close to the zero crossing of $\mathrm{d}\langle\phi_N\rangle/\mathrm{d}\rho_N$, which identifies the polarization maximum in the simulation data. The bifurcation density $\rho_c=18.1$ agrees well with the minimum of $\mathrm{d}\langle\phi_N\rangle/\mathrm{d}\rho_N$ on the simulation curve, where $\mathrm{d}^2\langle\phi_N\rangle/\mathrm{d}\rho_N^2=0$, indicating the most rapid decrease of $\phi_N$ with $\rho_N$ near $\rho_c$.

\section{$\langle\phi_N\rangle$ versus $\rho_N$ for different delays $\tau$}

\begin{figure}[htbp]
	\centering
	\includegraphics[width=0.6\linewidth]{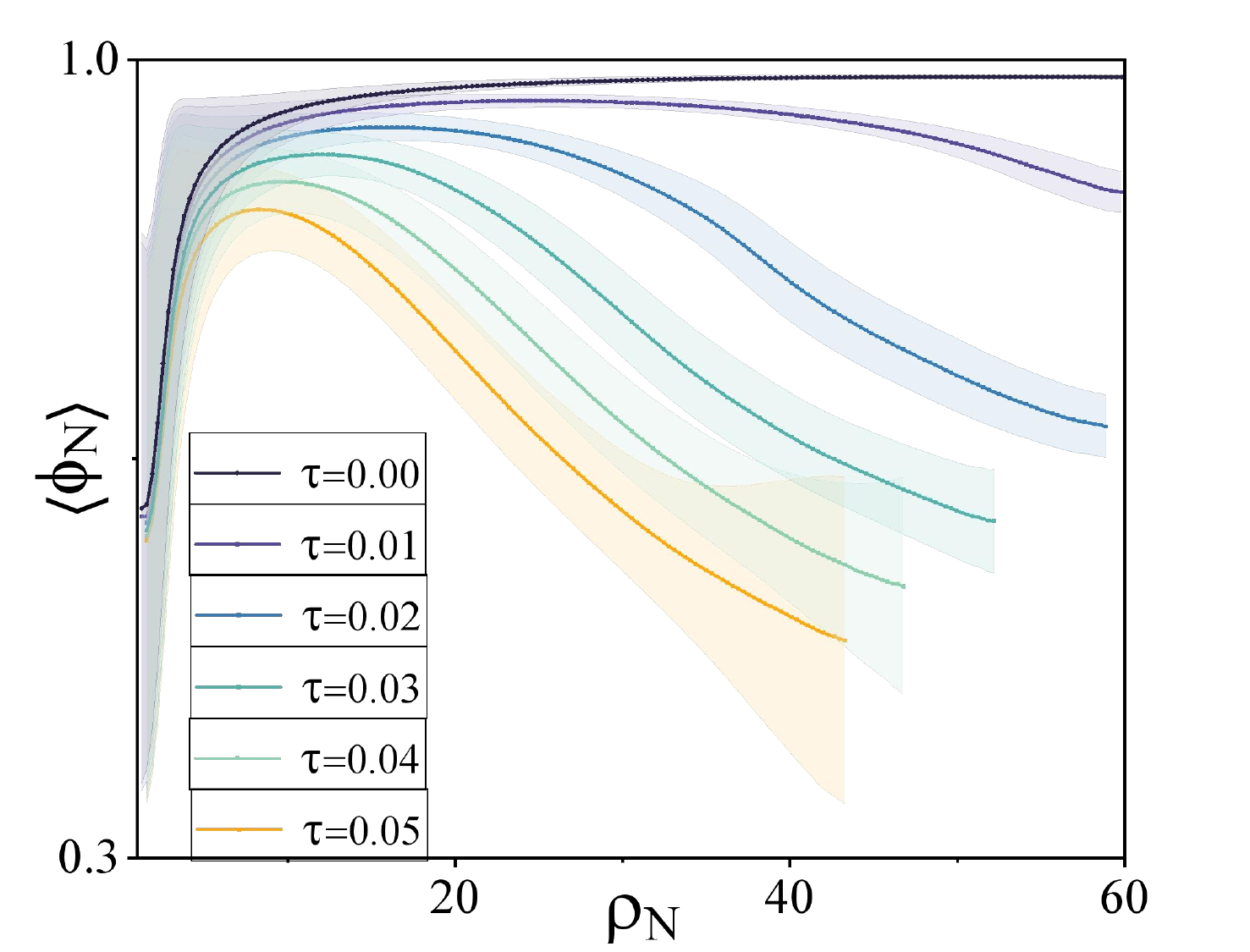}
	\caption{The average neighborhood polarization $\langle\phi_N\rangle$ versus the neighborhood density $\rho_N$ for different delays $\tau$}
	\label{fig:fig-Neighbor}
\end{figure}

In Fig.~\ref{fig:fig-Neighbor}, unlike the delay-free case, the average neighborhood polarization $\langle\phi_N\rangle$ for $\tau>0$ first increases and then decreases with increasing neighborhood density $\rho_N$. As the delay $\tau$ increases, $\langle\phi_N\rangle$ decreases globally, and its peak shifts to lower $\rho_N$, implying a more pronounced anomalous density--polarization relation (DPR). The $\langle\phi_N\rangle$--$\rho_N$ relation is qualitatively consistent with the $\langle\phi_L\rangle$--$\rho_L$ curves in the main text.

\section{Scaling collapse of theoretical polarization curves}

\begin{figure}[htbp]
	\centering
	\includegraphics[width=0.6\linewidth]{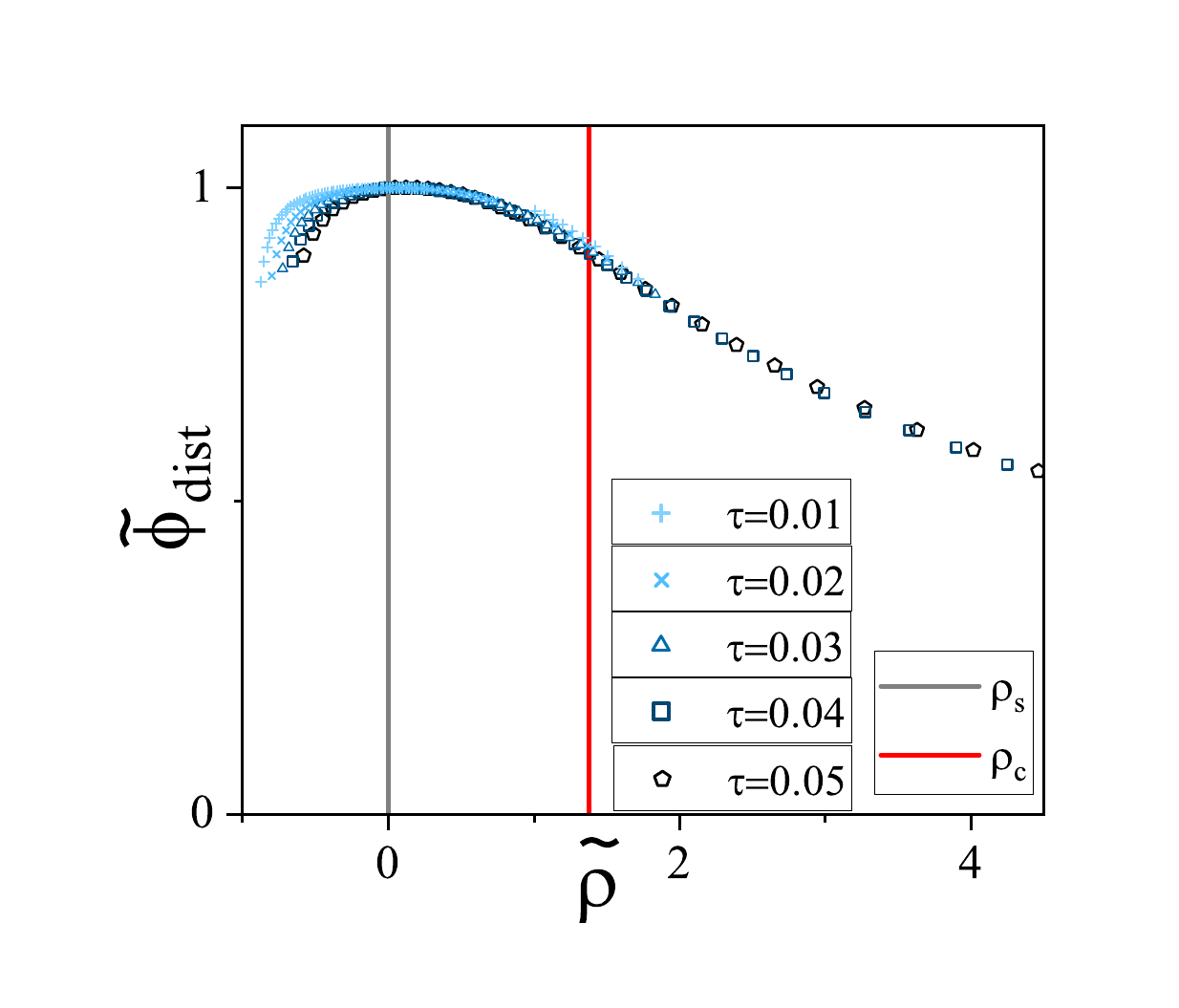}
	\caption{Theoretical scaled curves of $\tilde\phi_\mathrm{dist}$ versus $\tilde\rho$ for nonzero delays. The gray and red lines mark $\tilde\rho=0$ and $\tilde\rho=\rho_c/\rho_s-1$, respectively, using $\rho_s=7.6$ and $\rho_c=18.1$ for $\tau=0.05$.}
	\label{fig:fig-TheoScal}
\end{figure}

For the nonzero delays considered, the theoretical $\phi_\mathrm{dist}(\rho)$ curves approximately collapse onto a single master curve when expressed in the scaled coordinates ($\tilde{\rho} = \frac{\rho-\rho_s}{\rho_s}$, $\tilde\phi_\mathrm{dist} = \frac{\phi_\mathrm{dist}}{\phi_{\max}}$), as shown in Fig.~\ref{fig:fig-TheoScal}. 

\section{Markov property of the delay-embedded dynamics}

\subsection{Markov process in the $(n_\tau+1)$-dimensional delay-embedded space}
Let $\psi_k=\psi(k\Delta t)$ and $\tau=n_\tau\Delta t$. The Euler--Maruyama discretization of Eq.~\eqref{eq:SStateApprox2} reads
\begin{equation}
\psi_{k+1}=\psi_k-\beta\Delta t\sin\psi_{k-n_\tau}
+\sqrt{2\eta\Delta t}\,\xi_k,
\label{eq:embedded-scalar}
\end{equation}
where $\xi_k$ are independent standard Gaussian variables, as in Eq.~\eqref{eq:stoch-dde}. We augment the state with the preceding $n_\tau$ values of $\psi$ to form the $(n_\tau+1)$-dimensional state vector
\begin{equation}
\mathbf{x}_k=(\psi_k,\psi_{k-1},\ldots,\psi_{k-n_\tau})^{\mathsf T}.
\label{eq:delay-state}
\end{equation}
The dynamics of $\mathbf{x}_k$ can be written as
\begin{equation}
\begin{aligned}
\mathbf{x}_{k+1}&=\mathbf{x}_k+\Delta t\,\mathbf{F}(\mathbf{x}_k)
+\sqrt{2\eta\Delta t}\,\xi_k\mathbf{e}_0,\\
F_0(\mathbf{x}_k)&=-\beta\sin\psi_{k-n_\tau},\\
F_r(\mathbf{x}_k)&=\frac{\psi_{k-r+1}-\psi_{k-r}}{\Delta t},\qquad r=1,\ldots,n_\tau,
\end{aligned}
\label{eq:embedded-update}
\end{equation}
where $\mathbf{e}_0=(1,0,\ldots,0)^{\mathsf T}$ and $\mathbf{F}(\mathbf{x})=\bigl(F_0(\mathbf{x}),\ldots,F_{n_\tau}(\mathbf{x})\bigr)^{\mathsf T}
=\mathbb{E}[\mathbf{x}_{k+1}-\mathbf{x}_k\mid\mathbf{x}_k=\mathbf{x}]/\Delta t$ is the deterministic driving force.

The last $n_\tau$ components of $\mathbf{x}_{k+1}$ are deterministic shifts of the first $n_\tau$ components of $\mathbf{x}_k$. The first component depends on $\mathbf{x}_k$ and the new noise $\xi_k$, which is independent of the current and all previous states. Thus, the discretized reduced equation Eq.~\eqref{eq:embedded-scalar} defines a Markov process in the $(n_\tau+1)$-dimensional delay-embedded space.

\subsection{Numerical test of the Markov property in the delay-embedded space}
We set $\tau=0.05$ and $\Delta t=0.01$ and test the Markov property in delay-embedded spaces of different dimensions, focusing on the six-dimensional case.
We quantify history dependence in the embedded trajectories using conditional mutual information (CMI):
\begin{equation}
	\begin{aligned}
		I_m&:=I(\mathbf{x}_{k-1};\mathbf{x}_{k+1}\mid\mathbf{x}_k)\\
		&=I(\psi_{k-m};\psi_{k+1}\mid\mathbf{x}_k)\\
		&=\int p(\psi_{k-m},\psi_{k+1},\mathbf{x}_k)
		\ln\frac{p(\psi_{k-m},\psi_{k+1}\mid\mathbf{x}_k)}
		{p(\psi_{k-m}\mid\mathbf{x}_k)
			p(\psi_{k+1}\mid\mathbf{x}_k)}\mathrm{d}\psi_{k-m}\mathrm{d}\psi_{k+1}\mathrm{d}\mathbf{x}_k,
	\end{aligned}
	\label{eq:markov-cmi}
\end{equation}
where $\mathbf{x}_k=(\psi_k,\ldots,\psi_{k-m+1})^{\mathsf T}$ and $m$ is the embedding dimension. For $m=6$, the preceding analysis predicts that the embedded dynamics are Markovian.
For a Markov process, the past and future are conditionally independent given the present, so $I_m=0$. A positive $I_m$ indicates history dependence.
To assess whether the embedded trajectories are consistent with the Markov property, we generate Markov trajectories by resampling and compare their CMI with that of the embedded trajectories. During resampling, we randomly reassign the values preceding similar embedded states while retaining each embedded state and its subsequent value. The reference range spans the minimum and maximum CMI estimates from the resampled Markov trajectories. Because of finite-sample effects, the CMI estimates for the Markov trajectories fluctuate around zero.

\begin{figure}[htbp]
\centering
\includegraphics[width=\linewidth]{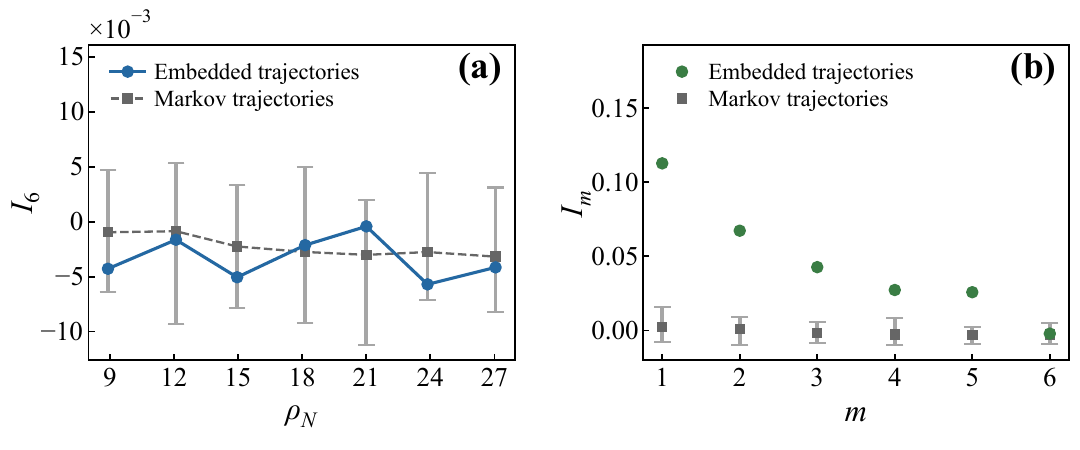}
\caption{CMI tests at $\tau=0.05$ and $\Delta t=0.01$. (a) Six-dimensional CMI $I_6$ versus density. (b) CMI $I_m$ versus embedding dimension $m$ near $\rho_N=\rho_c$. Circles and squares denote CMI estimates from embedded trajectories and locally resampled Markov trajectories, respectively. Gray error bars show the reference ranges.}
\label{fig:markov-cmi}
\end{figure}

As $\rho_N$ varies, the six-dimensional estimated CMI $I_6$ for the embedded trajectories remains close to zero and lies within the reference ranges of the Markov trajectories [Fig.~\ref{fig:markov-cmi}(a)]. Near $\rho_c$, the estimated CMI $I_m$ decreases with increasing embedding dimension $m$ and becomes comparable to that of the Markov trajectories at six dimensions [Fig.~\ref{fig:markov-cmi}(b)]. These results support the Markov property of the six-dimensional delay-embedded dynamics.

\section{Landscape--flux theory and Jacobian spectra}
\label{sec:landscape-flux}

\subsection{Landscape, flux, and force decomposition}

The nonequilibrium landscape-flux theory has been widely used to study nonequilibrium systems, with applications in many areas~\cite{WangPotential2008,WangLandscape2015,FangNonequilibrium2019,SuDynamical2024,YanMechanism2025,JianUncovering2025}. 

We consider a stochastic system whose state vector $\mathbf{x}$ evolves according to the following Langevin equation:
\begin{equation}
	\frac{\mathrm{d}\mathbf{x}}{\mathrm{d}t}=\mathbf{F}(\mathbf{x})+\boldsymbol{\zeta}(\mathbf{x},t),
	\label{LE}
\end{equation}
where $\mathbf{F}$ is the deterministic driving force and $\boldsymbol{\zeta}$ is the stochastic force with $\langle\boldsymbol{\zeta}(\mathbf{x},t)
\boldsymbol{\zeta}^{\mathsf T}(\mathbf{x},t^\prime)\rangle
=2\mathbf{D}(\mathbf{x})\delta(t-t^\prime)$, where $\mathbf{D}(\mathbf{x})$ is the diffusion tensor. For the delay-embedded dynamics, the state vector is $\mathbf{x}=(x_0,\ldots,x_{n_\tau})^{\mathsf T}$, with components $x_r=\psi(t-r\Delta t)$ for $r=0,\ldots,n_\tau$. Owing to the equivalence between the Langevin and Fokker--Planck descriptions, Eq.~\eqref{LE} can be rewritten as the Fokker--Planck equation for the probability distribution $P(\mathbf{x},t)$:
\begin{equation}
\frac{\partial P(\mathbf{x},t)}{\partial t}=-\nabla\cdot\mathbf{J}(\mathbf{x},t),
\end{equation}
where $\mathbf{J}$ is the nonequilibrium probabilistic flux. At the nonequilibrium steady state, the distribution becomes time independent, so that $\nabla\cdot\mathbf{J}_\mathrm{ss}=0$, where the subscript $ss$ denotes the steady state. In contrast to the vanishing flux at equilibrium, the nonequilibrium steady state only requires $\mathbf{J}_\mathrm{ss}$ to be divergence-free, and hence $\mathbf{J}_\mathrm{ss}$ can persist as a rotational field. From the Fokker--Planck equation, $\mathbf{J}_\mathrm{ss}$ reads:
\begin{equation}
	\mathbf{J}_\mathrm{ss}(\mathbf{x})=\mathbf{F}(\mathbf{x})P_\mathrm{ss}(\mathbf{x})-\nabla\cdot(\mathbf{D}P_\mathrm{ss}(\mathbf{x})).
	\label{Jss}
\end{equation}
In analogy with the equilibrium potential, the nonequilibrium potential landscape in the steady state is defined as
\begin{equation}
	U_\mathrm{neq}(\mathbf{x})=-\ln P_\mathrm{ss}(\mathbf{x}).
\end{equation}
Different from the equilibrium case, in which the driving force derives solely from the potential gradient, $\mathbf{F}$ can be decomposed into three parts:
\begin{equation}
	\mathbf{F}(\mathbf{x})=\frac{\mathbf{J}_\mathrm{ss}}{P_\mathrm{ss}}-\mathbf{D}\nabla U_\mathrm{neq}+\nabla\cdot\mathbf{D}
	=\mathbf{F}_\mathrm{curl}+\mathbf{F}_\mathrm{grad}+\mathbf{F}_D,
	\label{Fde}
\end{equation}
where $\mathbf{F}_\mathrm{curl}:=\mathbf{J}_\mathrm{ss}/P_\mathrm{ss}$ is the curl component of the driving force (a rotational flux force), $\mathbf{F}_\mathrm{grad}:=-\mathbf{D}\nabla U_\mathrm{neq}$ is the gradient force of the potential landscape, and $\mathbf{F}_D:=\nabla\cdot\mathbf{D}$ is the diffusion force arising from the spatially dependent noise. Note that the curl force equals the probability velocity field $\mathbf{v}_\mathrm{ss}$ (computed via central differences), i.e., $\mathbf{F}_\mathrm{curl}=\mathbf{v}_\mathrm{ss}:=\mathbb{E}\!\left[\frac{\mathbf{x}(t+\Delta t)-\mathbf{x}(t-\Delta t)}{2\Delta t}\,\middle|\,\mathbf{x}(t)=\mathbf{x}\right]$, so that $\mathbf{J}_\mathrm{ss}=P_\mathrm{ss}\mathbf{v}_\mathrm{ss}$. 
For notational simplicity, we henceforth drop the subscripts on $\mathbf{J}$, $\mathbf{v}$, $P$, and $U$ (the subscript $\mathrm{ss}$ for the first three and $\mathrm{neq}$ for $U$) in the following and in the main text.

In the main text, we take $\tau=0.05$ and $\Delta t=0.01$ as an example to show the anomalous behavior of the delayed CAP system. The corresponding delay-embedded state is $\mathbf{x}=(x_0,\ldots,x_5)^{\mathsf T}$, where $x_r=\psi(t-r\Delta t)$ for $r=0,\ldots,5$. To visualize the six-dimensional landscape and flux, we project onto the two-dimensional plane $\mathbf{y}=(x_5,x_0)^{\mathsf T}=(\psi(t-\tau),\psi(t))^{\mathsf T}$ and integrate over the remaining coordinates $\mathbf{z}=(x_1,x_2,x_3,x_4)^{\mathsf T}$. Hence, the projected landscape and flux are
\begin{equation}
\begin{aligned}
U_2(\mathbf{y})&=-\ln P_2(\mathbf{y}),
\qquad P_2(\mathbf{y})=\int P_6(\mathbf{y},\mathbf{z})\,\mathrm{d}\mathbf{z},\\
\mathbf{J}_2(\mathbf{y})&=\int
\begin{pmatrix}J_{6,5}(\mathbf{y},\mathbf{z})\\J_{6,0}(\mathbf{y},\mathbf{z})\end{pmatrix}\,\mathrm{d}\mathbf{z}.
\end{aligned}
\label{eq:projected-landscape-flux}
\end{equation}
Here, $P_6(\mathbf{x})$ and $\mathbf{J}_6(\mathbf{x})=(J_{6,0}(\mathbf{x}),\ldots,J_{6,5}(\mathbf{x}))^{\mathsf T}$ denote the probability density and flux in the six-dimensional delay-embedded space, respectively.

To quantify the global circulation of the flux, we define the mean flux as the spatial average of $|\mathbf{J}|$ over the delay-embedded domain $\Omega$:
\begin{equation}
	J_a:=\langle|\mathbf{J}|\rangle=\frac{1}{|\Omega|}\int_\Omega|\mathbf{J}|\,\mathrm{d}\mathbf{x}.
\end{equation}
The mean flux measures the average strength of the flux field. In the main text, $J_a$ is computed in the 6-dimensional delay-embedded space for each neighborhood density $\rho_N$, and its density slope $J_a':=\mathrm{d}\langle|\mathbf{J}|\rangle/\mathrm{d}\rho_N$ provides a dynamical signature of the bifurcation.

\subsection{Jacobian spectra and stability}
\label{sec:jacobian-origin}

For $n_\tau=5$, the linearized delay-difference equation~\eqref{eq:linear-dde} for perturbations about $\psi^*=0$ can be written in the delay-embedded space as
\begin{equation}
\mathbf{u}_{k+1}=\mathbf{u}_k+\Delta t\,A_F\mathbf{u}_k=M\mathbf{u}_k,
\label{eq:embedded-linear-map}
\end{equation}
where $\mathbf{u}_k=(u_k,u_{k-1},\ldots,u_{k-5})^{\mathsf T}$ and
\begin{equation}
M:=
\begin{pmatrix}
1&0&0&0&0&-\beta\Delta t\\
1&0&0&0&0&0\\
0&1&0&0&0&0\\
0&0&1&0&0&0\\
0&0&0&1&0&0\\
0&0&0&0&1&0
\end{pmatrix},
\qquad
A_F:=\operatorname{Jac}\mathbf{F}(\mathbf{0})=\frac{M-\mathbb{I}}{\Delta t}.
\label{eq:total-jacobian}
\end{equation}
Here, $\mathbb{I}$ denotes the $6\times6$ identity matrix.

Let $a_j$ denote the Jacobian eigenvalues of the driving force averaged over $1\Delta t$, and $\mu_j$ the eigenvalues of the one-step map $M$. In the corresponding continuous-time linear dynamics, the transient driving force has Jacobian $A_{\mathrm{inst}}:=\operatorname{Log}M/\Delta t$. Its eigenvalues $\lambda_j$ satisfy
\begin{equation}
\mu_j=1+\Delta t\,a_j=e^{\lambda_j\Delta t},
\qquad
\operatorname{Re}\lambda_j=\frac{\ln|\mu_j|}{\Delta t}.
\label{eq:three-spectra}
\end{equation}
At the bifurcation, $\max_j\operatorname{Re}\lambda_j$ crosses zero from below, indicating that the fixed point loses stability.

\begin{figure}[htbp]
\centering
\includegraphics[width=0.6\linewidth]{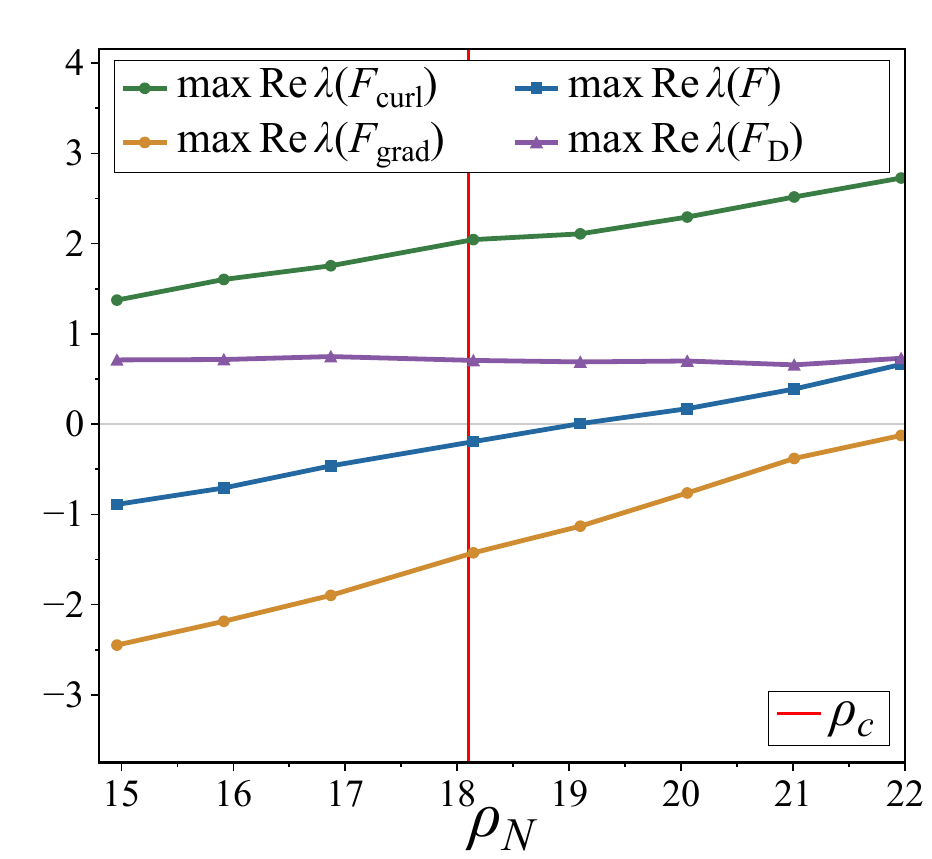}
\caption{Largest real parts of the Jacobian eigenvalues of the transient curl (green), gradient (orange), total (blue), and diffusion (purple) forces at the six-dimensional origin, for $\tau=0.05$ and $\Delta t=0.01$. The red line marks $\rho_c$.}
\label{fig:jacobian-stability}
\end{figure}

Applying the same construction to each component force in Eq.~\eqref{Fde}, we obtain the Jacobian eigenvalues $\lambda_{q,j}$ of the corresponding transient force, where $q\in\{\mathrm{curl},\mathrm{grad},D\}$. Near the bifurcation, $\max_j\operatorname{Re}\lambda_{q,j}$ remains positive for the curl force, negative for the gradient force, and slightly positive for the diffusion force, as shown in Fig.~\ref{fig:jacobian-stability}. Thus, the potential landscape tends to stabilize the fixed point, whereas the flux tends to destabilize it. Together, these results support the flux as the dynamical origin of the anomalous DPR.

\bibliographystyle{unsrt}
\bibliography{DCAP}